%% file: alamjporous2014.tex
\documentclass[onecolumn,final,12pt]{article}

\usepackage{epsfig} 
\usepackage{units,amsmath,amssymb,bm}
\usepackage[comma,numbers,compress]{natbib}

\usepackage{nomencl}
\makenomenclature

\title{Pressure drag and skin friction effects in a miscible flow through porous media}

\author{Jahrul M Alam\thanks{Correspondence author.}
  and M. Jalal Ahammad\thanks{Email: mja731@mun.ca} \\
      Department of Mathematics and Statistics\\
      Memorial University of Newfoundland\\
      St John's, NL\\
      A1C 5S7, Canada\\
      Email: alamj@mun.ca
}

\begin{document}

\maketitle    

\begin{abstract}

  Understanding miscible flow of CO$_2$ and oil in a porous media is an important topic in the field of petroleum engineering. In this article, we present an upscaling methodology for displacing oil by CO$_2$, and study the statistical mechanical theory to develop a subgrid scale model for the effect of CO$_2$-oil dissolution on the overall skin friction and pressure drag experienced by the porous media. We present a multiscale computational methodology for a near optimal calcuation of the velocity and pressure. We study the field scale sweep efficiency using a streamline based Lagrangian scheme. The simulation results show a good agreement with asymptotic analysis results.  
\end{abstract}

\nomenclature{CFD}{Computational fluid dynamics.}

\input{co2lag2013src}

\bibliographystyle{model1-num-names}
\bibliography{alamjco2lag2013}
\end{document}

%% file: co2lag2013src.tex
\section{Introduction}
\begin{figure}[htbp]
  \centering
  \includegraphics[width=9cm]{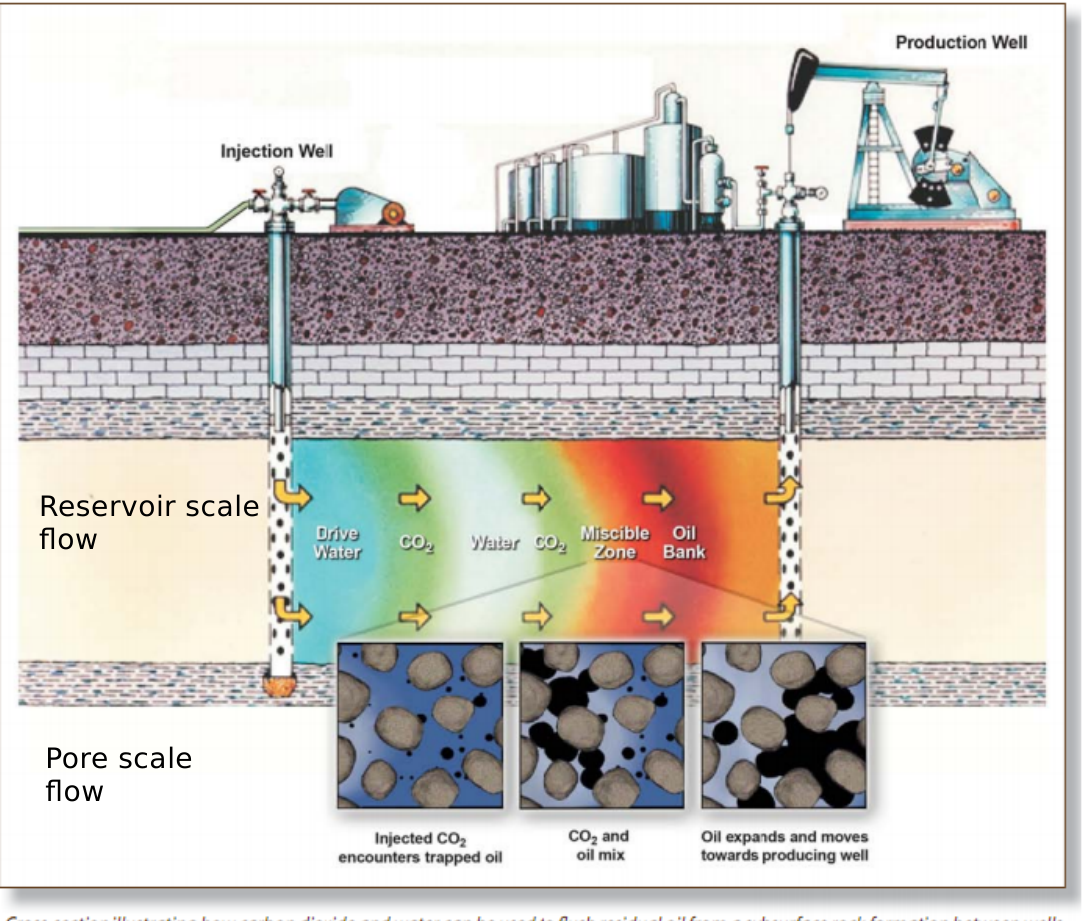}
  \caption{A schematic illustration of the miscible displacement of oil by a solvent in a vertical cross-section of an oil reservoir. A solvent is injected through an injection well, and oil is recovered through a production well. At the reservoir scale, shear stress on the walls affect the displacement efficiency. At the pore scale, the fluid interacts with random solid bodies. (Fig obtained from Google image.)}
  \label{fig:eor}
\end{figure}

The study of miscible displacement process in a porous medium -- also known as miscible flood or miscible drive -- is an important topic in the field of petroleum science and engineering~\cite{Zhao2012,Leetaru2009,Mohsin2009,Nobakht2007}. The miscible displacement of an oil by a solvent ({\em e.g.} carbon dioxide, CO$_2$) is a popular technique that is used for enhancing the recovery of the residual oil that is trapped by rocks or sands. The trapped oil is about $60$\% of the original oil in a reservoir, and is approximately $375$ billion barrels in the known oil fields of the United States~\cite{Sen2008}.
Since CO$_2$ is less viscous than oil, a binary mixture of CO$_2$ and oil experiences a reduced drag force when it flows through porous media. For example, while working on our car's engine, one notices that a solvent clears every trace of oil from tools, but water alone does not clean the tools. In view of the fluid dynamics, adding the solvent reduces the drag force, and increases the mobility of the fluid. The miscible flood process is more complex than what is demonstrated schematically in Fig~\ref{fig:eor}. There are two important topics for investigation -- one is how to enhance the mobility of oil at pore scale and the other is how to quantify the displacement efficiency -- the efficiency of sweeping out oil at the reservoir scale. Note the pore scale is $\mathcal O(\mu\unit{m})$ and the reservoir scale of $\mathcal O(100\unit{m})$. 
On the pore scale, the displacement of oil is affected by the spatial variation of reservoir permeability, oil swelling, or solvent composition and pressure. On the field scale, the displacement efficiency is affected by viscous fingering, solvent channeling, or gravity override. Computational modelling (also called reservoir simulation) often helps to study efficient oil recovery process.

In this article, we present a computational model for studying the fluid dynamics of a miscible flood, where a fluid ({\em e.g.} oil) is displaced by a solvent ({\em e.g.} CO$_2$) in a porous reservoir.
Our approach is based on the work of~\citet{Peaceman62} who verified such a mathematical model in comparison to another experimental model~\cite{Blackwell59}, where the effect of the drag force is assumed linearly proportional to the macroscopic velocity. 
%
Refs~\cite{Chen85,Homsy87,Tan88,Lenormand89,Birendra2011} studied viscous fingering, where the viscosity considered as a function of solvent concentration.
\citet{Gavin2004} presented experimental data, showing the departure from what was predicted by the commonly used mathematical model (see also, \cite{Rongze2012}). We have extended the miscible flow model of~\citet{Peaceman62} to simulate displacement efficiency, such as the fraction of oil that is swept from the unit volume of a reservoir upon injection of a solvent. The present article reports our primary results on modelling the time averaged total drag force
$$
f_i = \overbrace{-\frac{1}{\Delta V}\iint_{\partial S}\left[\frac{1}{T_0}\int_t^{t+T_0}\tilde Pn_idt'\right] dS}^{\hbox{pressure drag}} + \underbrace{\frac{\mu}{\Delta V}\iint_{\partial S}\left[\frac{1}{T_0}\int_t^{t+T_0}\frac{\partial\tilde u_i}{\partial n}dt'\right]dS}_{\hbox{skin friction}},
$$
per unit volume, which is due to the porous media. We decompose the force into two components, $f_i = f_i^D+f_i^s$, where $f_i^D$ is based on the Darcy's experiment~(see,~\cite{Peaceman62}). We have developed a mathematical model for the solvent induced component $f_i^s$ of the drag force. We are also interested in modelling multiphase flow in porous media using a computational fluid dynamics~(CFD) approach, where the Navier-Stokes equation is solved. The hope is to study a details of the transient behavior of the miscible displacement in porous media, where we employ a streamline based Lagrangian method to accurately resolve the sharp interface between two phases. 

Primary goals of this article include the following. First, we want to study an upscaling approach to resolve the small scale transient variability of a multiscale miscible flow in a porous media. Note, simply applying a high resolution numerical method may not capture such a small scale physics. Second, we study the statistical mechanical theory of viscosity to model the effect of CO$_2$-oil dissolution on the overall pressure drag and skin friction.  Third, we outline briefly a multilevel methodology for computing velocity and pressure with the generalized model at a near optimal computational cost. Here, we study how to retain a staggered arrangement between the velocity and pressure on a multilevel grid, and how to optimally address computational issues associated with modelling small scale transient variability of a miscible flow. Fourth, we show that sweep efficiency for displacing oil by CO$_2$ can be analyzed with sufficient accuracy using a streamline based Lagrangian model. All spatial derivatives have been approximated with second order finite difference schemes, where the time integration of the viscous force has been treated with a second order implicit scheme.   

This article presents the new development and a brief outline of the related material. Section~\ref{sec:ups} outlines the space-time intrinsic model for capturing small scale transient variability and its relationship with classical approaches, where a statistical mechanical approach is developed for resolving the solvent induced pressure drag and skin friction. Section~\ref{sec:cmp} summarizes the multilevel and Lagrangian methodologies. Section~\ref{sec:std} presents the results of numerical simulation. Finally, section~\ref{sec:dis} discusses the findings of the present research, as well as further extension of this development.

\begin{figure}[htbp]
  \centering
  \begin{tabular}{cc}
    \includegraphics[trim=4.5cm 11.5cm 4.0cm 11cm,clip=true,height=4cm]{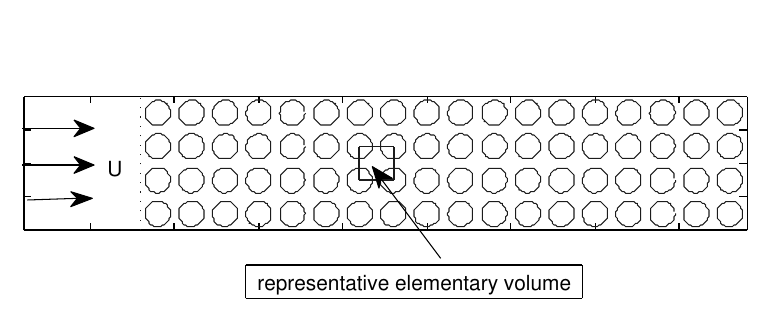}
    &
    \includegraphics[trim=4.0cm 9cm 2.5cm 8cm,clip=true,height=3cm]{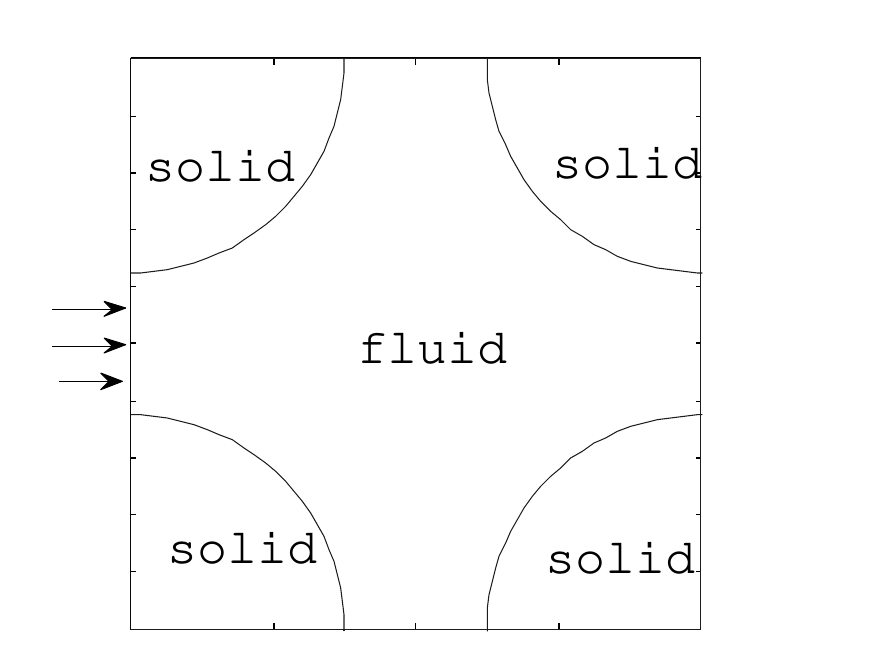}
    \\
    $(a)$&
    $(b)$
  \end{tabular}
  \caption{$(a)$ A permeable media enclosed by two impermeable boundaries, where a fluid enters with a known upstream velocity $U$; $(a)$~A representative elementary volume that contains both solid and fluid.}
  \label{fig:rev}
\end{figure}

\section{The computational fluid dynamics approach}\label{sec:ups}
The working fluid is assumed to have the property of crude oil, and is being displaced by a second fluid that is assumed to have the property of CO$_2$. The miscible displacement of oil and CO$_2$ occurs in an idealized reservoir, which is an isotropic porous media that consists of an array of rocks  -- a cross sectional view of which is shown in Fig.~\ref{fig:rev}($a$), and a representative elementary volume~(REV) of which is shown in Fig.~\ref{fig:rev}$(b)$. There are three characteristic length scales, $d$~($\unit{m}$), $\lambda~(\unit{m})$, and $H~(\unit{m})$, which represent the average pore space, size of the REV, and the height of the reservoir, respectively. The porosity is defined by $\phi = \Delta V_f/\Delta V$, where $\Delta V$ denotes the volume of the REV and $\Delta V_f$ denotes the volume of fluid (or void space) in the REV.

\subsection{The upscaling methodology}\label{sec:cua}

The upscaling methodology aims to model the `true' flow at scale $d\ll\lambda$ by an `approximate flow' at scale $\lambda < H$~\cite{Garibotti2009}. In other words, the simultaneous space-time mean 
$$
u_i^D = \frac{1}{\Delta V}\int_{\Delta V}\frac{1}{T_0}\int_t^{t+T_0}\tilde u_i(x_i,t')dt'dV
$$
of a pore scale quantity, $\tilde u_i$, can be considered as an approximation of the true flow; {\em i.e.} $u_i^D$ is an upscaling of the pore scale quantity $\tilde u_i$. Since $u_i^D$ does not vary in the volume $\Delta V$, the macroscopic momentum conservation law
$$
\frac{\partial}{\partial t}\int_{\Delta V}\rho u_i^Ddv +
\iint_{\partial S}\rho u_i^Du_j^Dn_jdS =
\iint_{\partial S}\left[-P\delta_{ij} +\tau_{ij} \right]n_jdS + \int_{\Delta V}f_idV,
$$
where $\tau_{ij} = \frac{\mu}{2}(\frac{\partial u_i^D}{\partial x_j} + \frac{\partial u_j^D}{\partial x_i})$, takes the form
\begin{equation}
  \label{eq:dc}
  \frac{\partial P}{\partial x_i} = -\frac{\mu}{K}u^D_i
\end{equation}
assuming that the total drag experienced by the porous media is given by $f_i = -\frac{\mu}{K}u^D_i$  in accordance with the result of Darcy's experiment on the flow of water through a sand column.
The Darcy's model~(\ref{eq:dc}) is an upscaling of the flow from the pore scale $(d)$ to a typically resolved scale $\lambda$, where $d\ll\lambda$, and states that the Reynolds number $\mathcal Re$ is related to the Darcy number $\mathcal Da$ by $\mathcal Re = \mathcal O(1/\mathcal Da)$. 

In the Darcy's model~(\ref{eq:dc}), effects of both inertia and viscous terms have been neglected. However, \citet{Bear72} suggested that the onset of inertial effect begins about at $Re_d$ between $3$ to $10$, which is confirmed by a number of experiments and numerical simulations~\cite{Zeng2006,Fancher33,Green51,Ergun52,Scheidegger74,Blick88}, although some experiments~\cite{Fancher33} observed the inertial effect at $Re_d < 3$. To gain an understanding on the range of $\mathcal Re_d$, \citet{Meyer85} examined $34$ consolidated and unconsolidated materials, and reported that $0.03\le\mathcal Re_d\le 1,000$, which explains the regime for the onset of inertial effect. In addition, deviation from the Darcy's model~(\ref{eq:dc}) was reported by many authors, for example, see the works of~\cite{Gavin2004}, \cite{Fand87}, \cite{Hubbert57}, \cite{Rongze2012}, and~\cite{Srinivasan2013}. The above analysis indicates that the effects of inertia and viscous stress are dominant at the length scale $<\lambda$; however, the overall momentum transfer law may be approximated with good accuracy by~(\ref{eq:dc}) at the length scale $\ge\lambda$.

In applications, $\lambda$ may be characterized by the grid resolution $\Delta$, and  when $H$ is  large, such as oil reservoir simulations,  there is a propensity to adopt a larger $\lambda\sim\Delta$, which is mainly constrained by the threshold on the number of computational degrees of freedom, {\em i.e.} $\mathcal N=(H/\lambda)^d$ in a $d$-dimensional model. Since large errors are also observed, advanced numerical methods~\cite{Jenny2006} are proposed to solve~(\ref{eq:dc}) at sufficiently high resolution. However, a high resolution simulation violates the fundamental assumption behind the Darcy's model that the characteristic length scale $\lambda$ should be large enough so that effects of inertia and viscous dissipation are not felt~\cite{Brinkman49,Ma93,Whitaker96}. A CFD model would be the most appropriate when high resolution simulation of a reservoir flow is desired.

\begin{figure}
  \centering
    \includegraphics[trim=4.5cm 8cm 4.0cm 8cm,clip=true,height=5cm]{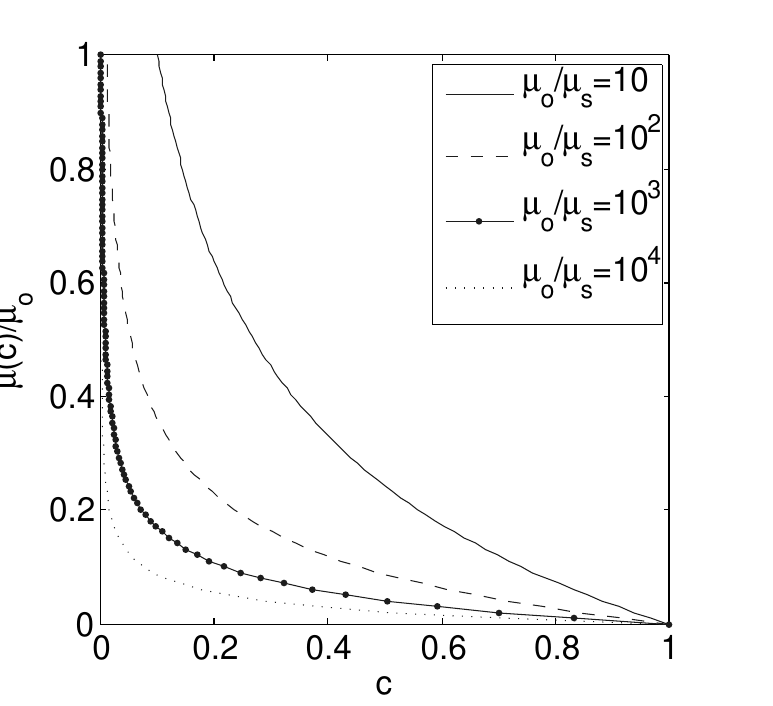}
  \caption{Normalized viscosity of a mixture of solvent and oil, $\mu(c)/\mu_o$, where  $\mu(c) = [c\mu_s^{-1/4} +(1-c)\mu_o^{-1/4}]^{-4}$, for several values of the mobility ratio, $\mu_o/\mu_s$.}
  \label{fig:muc}
\end{figure}

Medium heterogeneity appears in~(\ref{eq:dc}) {\em via} spatially dependent anisotropic porosity $K$, a resolution of which is a principal motivation behind the pioneering work of~\cite{Jenny2006}, and is not the primary focus of the present article. For a miscible flow of two fluids, modelling the effect of reduced pressure drag and skin friction -- as also suggested by~(\ref{eq:dc}) when the viscosity~($\mu$) is reduced by the addition of a solvent -- is a primary contribution of the present article. Two empirical models are available -- one is $\mu(c) = [c\mu_s^{-1/\gamma} +(1-c)\mu_o^{-1/\gamma}]^{-\gamma}$, where $c$ is the solvent concentration, and  $\gamma$ is either $4$ or $=-1$~\cite{Koval63}, and the other is $\mu(c) = \exp(c\ln(\mu_o/\mu_s)$~\cite{Sahu2009}. For $\gamma=4$, Fig.~\ref{fig:muc} shows the nonlinear dependence $\mu(c)$ for various values of the mobility ratio $\mu_o/\mu_s$. With a high mobility ratio, a small fraction of the solvent leads to a large reduction of viscosity, which means that the drag of the porous media on a miscible flow is reduced with the concentration. 
\subsection{A brief outline of the present CFD model}
\citet{Peaceman62} proposed the upscaling model, eq~(\ref{eq:dc}, \ref{eq:inc2d},~\ref{eq:mnc}), for a miscible flow through a porous media of permeability $K$~($\unit{m^2})$ and porosity $\phi$~(\%), for which, the search for the best approximation on the concentration dependent viscosity $\mu(c)$ remains an active research topic~\cite{Homsy87,Birendra2011,Birendra2013}.
%
%

%
First, we employ a simultaneous space-time {\em intrinsic} mean of the momentum conservation law, in order to model fluctuations with respect to $u_i^D$ at two scales. Second, we study a statistical mechanical approach to model the dependence of viscosity on the solvent concentration, $\mu(c)$.

Let the velocity, $\tilde u_i$, (or another variable) be decomposed as
$
\tilde u_i = u^D_i + u'_i + u''_i,
$
where $u'_i$ and $u''_i$ are deviations associated with length scales $\lambda$ and $d$, respectively, where  the space-time mean of $u'_i$ and $u''_i$ -- as defined in Sec~\ref{sec:cua} -- are zero. Let us define the intrinsic average in space and time (see,~\cite{DeLemos2006})
$$
\langle u_i\rangle\equiv u_i = \frac{1}{\Delta V^f}\int_{\Delta V}\frac{1}{T_0}\int_t^{t+T_0}\tilde u_i(x_i,t')dt'dV,
$$
which assumes that $u_i = u_i^D+u'_i$ and the intrinsic average of only $u''_i$ vanishes. We have dropped $\langle\cdot\rangle$ for simplicity. In other words, the missing details in the space-time average, $u_i^D$, has been recovered or retained in the intrinsic representation $u_i$. Furthermore, the space-time average ($u_i^D$) and the intrinsic average ($u_i$) are related by $u_i^D = \phi u_i$, which states that the space-time averaged flow resolves only a fraction of the intrinsic flow. In the space-time average $u_i^D$, local variation from the interaction between the fluid and the porous media is ignored, which has been modelled in the intrinsic representation. This assumption is important in miscible flow because the mixing occurs at a length scale that is much smaller than $d$. The technical details of the averaging process -- also known as the double decomposition -- are documented by~\cite{Bear72}, \cite{Whitaker99}, \cite{Lage2002}, and~\cite{DeLemos2006}, where the space and time average are applied sequentially to model turbulent flows in the porous media. For the laminar flow considered in the present work, the averaging process has been adapted  to resolve $u'_i$ through the intrinsic average, and to model the hydrodynamic dispersion associated with the unresolved component $u''_i$. 

If we take the intrinsic space-time average of the momentum conservation law in the differential form, the divergence of the flux of momentum (divided by $\rho$) for a constant density fluid takes the form
$$
\left\langle\frac{\partial\tilde u_i\tilde u_j}{\partial x_j}\right\rangle 
= \frac{\partial u_iu_j}{\partial x_j} + 
\frac{\partial\langle u''_iu''_j\rangle}{\partial x_j} + 
\left[\frac{\partial\tau''_{ij}}{\partial x_j} \right],
$$
where $\langle\cdot\rangle$ is used to denote the intrinsic space-time average, and $\langle\tilde u_i\tilde u_j\rangle - u_iu_j - \langle u''_iu''_j\rangle = \tau''_{ij}$. In fact, the term in $[\cdots]$ would represent both the Reynolds stress and the turbulent dispersion~\cite{Lage2002}, which are not considered in the present work, and hence, $\tau''_{ij}$ will be dropped. The other term $\langle u''_i u''_j\rangle$ represents dispersion associated with spatial fluctuation of pore scale time mean velocity, which may be modelled by $\langle u''_iu''_j\rangle\sim\mu_{\hbox{\tiny eff}}\frac{\partial u_i}{\partial x_j}$~\cite{Brinkman49}.

Similarly, the intrinsic space-time averages of the pressure term, $-\tilde P\delta_{ij}$, and the deviatoric stress, $\tau_{ij}$, provide two additional terms representing the pressure drag and the skin friction, {\em i.e.}
$$
f_i = \overbrace{-\frac{1}{\Delta V}\iint_{\partial S}\left[\frac{1}{T_0}\int_t^{t+T_0}\tilde Pn_idt'\right] dS}^{\hbox{pressure drag}} + \underbrace{\frac{\mu}{\Delta V}\iint_{\partial S}\left[\frac{1}{T_0}\int_t^{t+T_0}\frac{\partial\tilde u_i}{\partial n}dt'\right]dS}_{\hbox{skin friction}},
$$
where we have assumed that the flow is incompressible and Newtonian.

\subsection{The intrinsic CFD model}
Applying the space-time intrinsic average to the first principle conservation laws of mass and momentum for a miscible flow in a porous media, the governing equations -- after normalizing by a characteristic velocity $U$ and a length scale $H$ -- takes the form
\begin{equation}
  \label{eq:inc2d}
\frac{\partial u_i}{\partial x_i}  = 0,
\end{equation}
\begin{equation}
  \label{eq:hrmm}
  \frac{\partial u_i}{\partial t} + u_j\frac{\partial u_i}{\partial x_j} = - \frac{\partial P}{\partial x_i} +\frac{\alpha}{\mathcal Re}\frac{\partial}{\partial x_j}\frac{\partial u_i}{\partial x_j} +f_i, 
\end{equation}
and
\begin{equation}
  \label{eq:mnc}
\frac{\partial c}{\partial t} + u_j\frac{\partial c}{\partial x_i}  = \frac{1}{\mathcal Re\mathcal Sc}\frac{\partial}{\partial x_j}\frac{\partial c}{\partial x_j},
\end{equation}
where the same symbol has been retained for dimensional and dimensionless quantities. 
Eqs.~(\ref{eq:inc2d}-\ref{eq:mnc}) include four dimensionless parameters: the Reynolds number $\mathcal Re = \frac{UH}{\nu}$, the Darcy number $\mathcal Da = \frac{K}{H^2}$, the Schmidt number $\mathcal Sc=\frac{\nu}{D}$, and the viscosity ratio, $\alpha = \frac{\mu_{\hbox{\tiny eff}}}{\mu_o}$. 
Eq~(\ref{eq:hrmm}) is a model of the momentum transfer law for a solvent-oil system in a porous media, where on the right hand side, the second last term appears as a model for the effect of the hydrodynamic dispersion $\langle u''_iu''_j\rangle\sim\mu_{\hbox{\tiny eff}}\frac{\partial u_i}{\partial x_j}$ associated with the unresolved component $u''_i$, and the last term $f_i$ accounts for the pressure drag and skin friction, appears as a result of the intrinsic average, and has to be modelled. Eq~(\ref{eq:mnc}) is the Peacemann model for miscible flooding of oil by a solvent~\cite{Peaceman62}, where $c(x_i,t)$ is the concentration of the solvent in a REV. Note also that if the terms on the left side of~(\ref{eq:hrmm}) are dropped, then~(\ref{eq:hrmm}) takes the form of Darcy-Brinkman-Forchheimer model for a porous media flow.

In a CO$_2$-oil system, the oil will be saturated with the dissolved CO$_2$, and there is an interface between the free CO$_2$ and CO$_2$ saturated oil. Thus, to model the drag force, $f_i$, we consider that the solvent induced mixture viscosity is given by $\mu(c) = \mu_o(1+\mu^c(x_1,x_2))$,  where $\mu^c(x_1,x_2)$ is due to the dissolved CO$_2$. Next, we assume that 
\begin{equation}
  \label{eq:frc}
  f_i= \underbrace{-\frac{\mu_o}{K} u^D_i}_{f_i^D} + f_i^s,
\end{equation}
and employ the viscosity theory of binary fluids to model the component, $f_i^s$.

\subsection{Viscosity of binary liquids}
There have been a number of approaches which can be used  to estimate the viscosity of a binary liquid~\cite{Dabir2007,Barus}.
In a molecular model, a liquid is a lattice of densely populated molecules, which assumes the functional dependence of viscosity and diffusion on other thermodynamic properties of the liquid~\cite{Eyring41}.
In such a model, the activated collision of molecules from one equilibrium state to another is an elementary process for transport mechanisms~\cite{Eyring36}. Since forces between closely packed molecules cause momentum transfer in response to a shearing stress, molecules have to overcome a potential barrier imposed by their neighbors~\cite{Brush62}. Thus, in a viscous flow,  the initial state of the molecules requires necessary energy -- known as the activation energy, $\Delta E$ -- to pass over the potential barrier before it reaches the final state~\cite{Eyring36,Eyring37,Eyring41}. Further insight into the role of activation energy on the viscosity of binary liquids can be found from the work of~\cite{Babu2011}, \cite{Dabir2007}, and~\cite{Bearman60}.

A solvent increases the volume flow rate~($Q$) of the solvent-oil mixture. Thus, the total drag force (pressure drag and skin friction) experienced by a porous media may be expressed as $f_i=f_i^D+f_i^s$, where
$f_i^D$ represents the drag as if no solvent is added, and 
$$
f_i^s =\rho_s\frac{\Delta E}{h} \frac{Q_i}{K},
$$
is the drag due to the fractional density $\rho_s$ of the solvent, where $h~(\unit{m^2{\cdot}kg{\cdot}s^{-1}})$ is the Plank constant. For a Newtonian flow of a solvent-oil model, the activation energy, $\Delta E$, is related to the potential barrier, $\mu^c$, in a way that the ratio of the shearing stress to the rate of deformation is proportional to $\mu^c$. Thus, the potential barrier, $\mu^c$, has the units of dynamic viscosity~($\unit{kg\cdot m^{-1}\cdot s^{-1}}$). Note that $\mu^c$ is a dynamic viscosity that is `induced' by the presence of solvent molecules, $\rho_s$.  Hence, in a volume $\mathcal V$ that is enclosed by the surface $\mathcal S$, the flux of gradient of the potential barrier ({\em i.e.} `induced' viscosity $\mu^c$) across $\mathcal S$ can be related to the probability density of solvent molecules, $\rho_s\Delta E/h$, through the Maxwell's theory of viscosity~\cite{Barus}
$$
\iint\limits_{\mathcal S}\frac{\partial\mu^c}{\partial x_i} dS_i = \iiint\limits_V\frac{\Delta E}{h}\rho_s dV,
$$
which leads to the Poisson equation
$$
\frac{\partial}{\partial x_j}\left(\frac{\partial\mu^c}{\partial x_j}\right) = \frac{\Delta E}{h}\rho_s.
$$
In a mixture, the probability that the solvent molecules will be in a given volume may follow a Gibbs distribution. Hence, if $\tilde\rho_s$ denotes the mixture density, we get
$$
\rho_s = \tilde\rho_s\left[1-\exp\left(-\frac{\Delta E}{\mathcal B\Theta}\right)\right],
$$
where $\mathcal B~(\unit{m^2{\cdot}kg{\cdot}s^{-2}{\cdot}K^{-1}})$ is the Boltzmann constant and $\Theta~(\unit{K})$ is the absolute temperature. We now proceed to establish the dependence of the necessary activation energy on three parameters, namely, the resistance due to the potential barrier, $\mu^c$, the medium permeability, $K$,  and the molar velocity, $v_m$, such that 
$$
\Delta E = g(K,v_m,\mu^c),
$$
where $g$ is an arbitrary function to be determined. A dimensional analysis leads to %
$$
\Delta E = Kv_m\mu^c,
$$
and in the limit of $\frac{Kv_m\mu^c}{B\Theta}\ll 1$, we get
\begin{equation}
  \label{eq:muc}
\frac{\partial}{\partial x_j}\left(\frac{\partial\mu^c}{\partial x_j}\right) = \frac{\mu^c}{\lambda_c^2},
\end{equation}
where the dimensionless quantity $\lambda_c$ is given by
$$
\lambda_c^2 = \frac{hB\Theta}{\Delta E\tilde\rho_s K v_m}.
$$
In order to have $\mu(c)=0$ on the impermeable walls, we have adopted
$$
\mu^c(x_1,y_{\min})=-1=\mu^c(x_1,y_{\max}),
$$
and to ensure the mass conservation,
$$
\frac{\partial\mu^c}{\partial x_1}(x_{\min},y)=0,\frac{\partial\mu^c}{\partial x_1}(x_{\max},y)=0.
$$
Putting the above calculations together, the drag force per unit volume -- in dimensionless variables -- is given by
\begin{equation}
  \label{eq:drgs}
  f_i = -\frac{\phi u_i}{\mathcal Re\mathcal Da} + \frac{C_{\varphi}}{\lambda_c^2\mathcal Re\mathcal Da}\mu^c\frac{\partial\varphi}{\partial x_i},
\end{equation}
where $Q_i$, representing a steady state, incompressible, and irrotational component of the velocity, has been written in terms of a velocity potential, $\varphi$, and $C_{\varphi}$ is an arbitrary constant. The velocity potential, $\varphi$,  for an incompressible flow is computed from
$$
\frac{\partial}{\partial x_j}\left(\frac{\partial\varphi}{\partial x_j}\right) = 0,
$$
$$
\varphi(x_{\min},x_2) = 1, \varphi(x_{\max},x_2) = 0, \frac{\partial\varphi}{\partial x_2}(x_1,y_{\min})=0=\frac{\partial\varphi}{\partial x_2}(x_1,y_{\max}).
$$
We have developed a multilevel methodology for solving the generalized upscaling model~(\ref{eq:inc2d}-\ref{eq:mnc}). 
\section{The computational methodology}\label{sec:cmp}
Two approaches have been employed to address the multiscale challenge of reservoir flows. First, the multilevel algorithm for the pressure and the incompressible velocity makes the proposed model asymptotically optimal, where the computational work load increases approximately linearly with the number of grid points, if the resolution increases for adapting the solution to the multiscale features. Second, the streamline based Lagrangian scheme resolves the associate multi-physics features of the miscible flow using a near optimal computational effort.  

\subsection{Discretization}

All spatial derivatives, including the advection terms, have been treated with a second order finite difference scheme. Viscous terms are integrated with a second order Crank-Nicolson scheme. For the time integration of all other terms, we have implemented three explicit schemes -- the Euler's method, the predictor-corrector method, and the Runge-Kutta method~\cite{Tannehill97}. In the following time integration methodology, computed $u_i$ and $c$ are assigned to initially known $u^n_i$ and $c^n_i$ for marching in time at each time step. 

A time step begins by computing the pressure associated with the incompressible velocity field from
\begin{equation}
  \label{eq:p}
  \frac{\partial^2}{\partial x_j\partial x_j}
  \left(\underbrace{P+\frac{u^n_iu^n_i}{2}}_{p'}\right)
  =
\frac{\partial}{\partial x_i}\left( f_i
  - u^n_j\left(\frac{\partial u^n_i}{\partial u^n_j} - \frac{\partial u^n_j}{\partial u^n_i} \right)
\right)
\end{equation}
using the multilevel methodology developed by~\cite{Alam2002}. We have noticed that the rotational form of the advection term is more accurate for conserving mass with the present multilevel algorithm. The required computational complexity of this algorithm is linearly proportional to the number of the grid points~\cite{Wesseling2000}. Using the computed $p'$ from~(\ref{eq:p}) and employing the same multilevel algorithm, the incompressible velocity field, $u_i$, is obtained from
\begin{eqnarray}
  \label{eq:cn}
  \nonumber
  \frac{2u_i}{\Delta t} -\frac{1}{\mathcal Re}\frac{\partial^2u_i}{\partial x_j\partial x_j}
  &&=
  \frac{2u_i^n}{\Delta t} +\frac{1}{\mathcal Re}\frac{\partial^2u_i^n}{\partial x_j\partial x_j}
  \\\nonumber
  &&
  -2\left[\frac{\partial}{\partial x_i}\left(\underbrace{P+\frac{u^n_iu^n_i}{2}}_{p'}\right)
  + u^n_j\left(\frac{\partial u^n_i}{\partial x_j}-\frac{\partial u^n_j}{\partial x_i}\right) \right. 
  \\
  &&\left.-\frac{\phi u^n_i}{\mathcal Re\mathcal Da}
  +\frac{C_{\varphi}}{\lambda_c^2\mathcal Re\mathcal De}\mu^c(x_1,x_2)\frac{\partial\varphi}{\partial x_i} 
  \right].
\end{eqnarray}
The concentration field, $c$, is computed {\em via} a fractional step algorithm. In the first step,
\begin{equation}
  \label{eq:c}
  \frac{2c^*}{\Delta t} -\frac{1}{\mathcal Re\mathcal Sc}\frac{\partial}{\partial x_j}\left(\frac{\partial c}{\partial x_j}\right)
  =
  \frac{2c^n}{\Delta t} +\frac{1}{\mathcal Re\mathcal Sc}\frac{\partial }{\partial x_j}\left(\frac{\partial c^n}{\partial x_j}\right)
\end{equation}
is solved for $c^*$ using the multilevel algorithm. In the second step, the streamline based Lagrangian scheme of~\cite{Alam2012b} is adapted to propagate $c^*$ along streamlines for updating $c$ at each grid point. Note that (\ref{eq:c})~is a discretization of~(\ref{eq:mnc}) ignoring the advection terms. The streamline methodology models the neglected advection terms. 
\subsection{A streamline based Lagrangian methodology}
The computed velocity field $u_i$ is tangential to a family of curves -- known as streamlines, which can be parameterized by $s_i(\xi)$ such that
\begin{equation}
  \label{eq:xi}
  \frac{\partial s_i}{\partial\xi} = u_i.
\end{equation}
Since the directional derivative for any curve $s_i$ in the direction of a vector field $u_i$ leads to $u_i = u_j\frac{\partial s_i}{\partial x_j}$, eq.~(\ref{eq:xi}) gives $\frac{\partial c}{\partial \xi}=u_j\frac{\partial c}{\partial x_j}$. As a result, the advection part of~(\ref{eq:mnc}) takes the form
\begin{equation}
  \label{eq:cxi}
  \frac{\partial c}{\partial t} + \frac{\partial c}{\partial\xi}=0.
\end{equation}
Therefore, a multidimensional advection equation is reduced to a one-dimensional linear advection equation, which can be solved analytically. The computational challenge of solving an advection equation is replaced with that of solving ODEs and interpolation, which removes the numerical artifacts associated with the solution of an advection equation.    
During this second fraction of the time step, the solution of~(\ref{eq:c}) is considered as the initial condition for~(\ref{eq:cxi}), and its solution is obtained analytically such that $c(\xi,t) = c^*(\xi-t)$. Since every value of $\xi$ corresponds to a point on the curve $s_i(\xi)$, the quantity $c(\xi,t)$ is always situated on a streamline, but may not necessarily be on a grid point. Thus the solution $c(\xi,t)$ of~(\ref{eq:cxi}) is interpolated onto grid points using the mass conserving algorithm presented by~\cite{Alam2012b}.

\section{Numerical simulation and discussion}\label{sec:std}
Assuming that miscibility between CO$_2$ and oil occurs only at the interface, and there is no background dissolution of CO$_2$, the effect of viscous stress, permeability, and the drag force of the porous media on the miscible displacement of CO$_2$ and oil has been studied. A number of numerical simulations have been performed with $\mathcal Re\ge 1$, $10^{-4}<\mathcal Da< 10^6$, $C_{\varphi}=0$, as well as $C_{\varphi}>0$. Note that $\mathcal Da=1$ represents a permeability that allows a flow rate $1~\unit{m^3/s}$ for a fluid at viscosity $1~\unit{Pa{\cdot}s}$ in the $1~\unit{m}$ long reservoir with a cross section of $1~\unit{m^2}$ if the pressure difference is $1~\unit{dyne/m^2}$. For a fixed $\mathcal Re$, the drag force weakens if $\mathcal Da >1$. The porosity $18\%$ is determined based on the representative elementary volume so that the present idealized model resembles an actual reservoir as closely as possible~\cite{Gozalpour2005,Teresa2013}. Other simulation parameters have been listed in Table~\ref{tab:frst}. Assuming a initially stationary situation, a mass of CO$_2$ is emplaced in the injection well near $x_1=0$, where the concentration, $c(x_1,x_2)$, has a Gaussian distribution in $x_1$ with no variation in the $x_2$ direction. A narrow strip of CO$_2$ followed by oil is considered because the overall drag is assumed proportional to $\mu_o$. In Fig.~\ref{fig:nuv1}$a$, the initial emplacement of CO$_2$ is marked red, which has been pushed at a constant rate of injection toward the production well that is located at $x_2=3$.

%
\begin{table}
\centering
 \begin{tabular}{|l|l|l}
 \hline 
 Parameter & value\\
 \hline
$L_1 \times L_2 $  & $3 \times 1$  \\  
$n_1 \times n_2 $ & $512 \times 128$  \\ 
$\Delta t$ & $10^{-2}$ \\
$\bm\nabla P_0$ & 2 \\
$\alpha$ & 1  \\
$Re Sc$ & 20000 \\
$\phi$ & 18\%\\
 \hline  
\end{tabular}
\caption{List of parameters for simulations with Reynolds number, $\mathcal Re\ge 1$ and Darcy number $10^{-4}<\mathcal Da< 10^6$.}
\label{tab:frst}
\end{table}

\subsection{Effect of viscous stress and permeability on miscible displacement}

To investigate the effect of viscous stress on the dynamics of a miscible displacement process, we have simulated the time evolution of CO$_2$ using $\mathcal Da=\phi/\mathcal Re$, $\alpha=1$, and $\mathcal Re\ge 1$, where the ratio of the momentum sink to the pressure gradient force is $\mathcal O(\frac{\phi}{\mathcal Re\mathcal Da})$. In other words, the Darcy's balance law~(\ref{eq:dc}) is satisfied. Here, $\mathcal Re=1$ characterize a reservoir of $1\,000~\unit{m}$ deep, where a crude oil of kinematic viscosity $3.5\times 10^{-3}~\unit{m^2{\cdot}s^{-1}}$  flows at a typical intrinsic velocity, $3.5\times 10^{-6}~\unit{m/s}$~\cite{Afsharpoor2012}. The CO$_2$ distributions at $\mathcal Re=1$ and $\mathcal Re=10^2$ have been presented in Fig.~\ref{fig:nuv1}$b$ and~\ref{fig:nuv1}$c$, respectively, for $\mathcal Da=\phi/\mathcal Re$. CO$_2$ has reached near $x_1=2$ along the centerline of the reservoir at $t=22.5H/U$ in Fig.~\ref{fig:nuv1}$b$ and at $t=2.3H/U$  in Fig.~\ref{fig:nuv1}$c$.
Note that $\mathcal Re=10^2$ represents that the strength of viscous stress is reduced by a factor of $100$ compared to a flow at $\mathcal Re=1$. Clearly, the distortion of the initial CO$_2$ is strongly influenced by the viscous stress. As explained by~\citet{Brinkman49}, the influence of viscous shearing stress near an isolated solid body embedded in a porous matrix is different than how the effect of viscous stress is felt on the the porous matrix itself. For a qualitative measure of the distortion of initial shape of the CO$_2$ sample, a rectangle has been drawn in each of the plots in Fig.~\ref{fig:nuv1}. We see that the proportion of CO$_2$ within the marked region has been increased at $\mathcal Re=10^2$ compared to $\mathcal Re=1$, showing that field scale sweep efficiency would improve if the shear stress weakens. 

\subsection{Effect of permeability}
\begin{figure}
  \centering
  \begin{tabular}{cc}
    \multicolumn{2}{c}{
      \includegraphics[width=5cm]{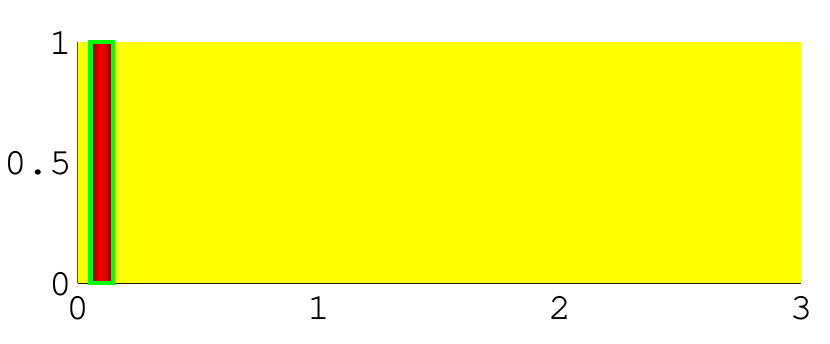}
    }\\
    \multicolumn{2}{c}{($a$)}\\
    \includegraphics[width=5cm]{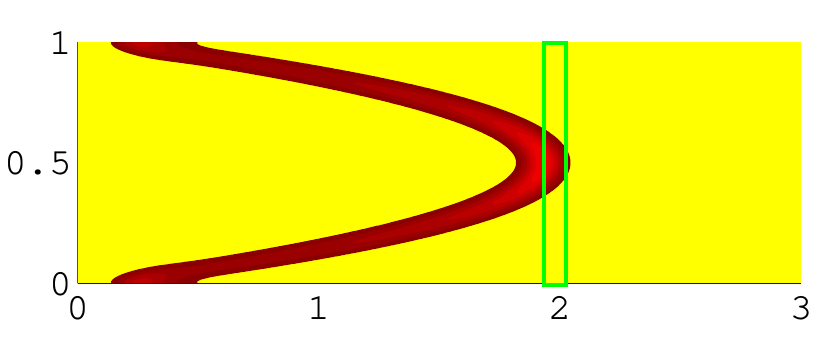}&
    \includegraphics[width=5cm]{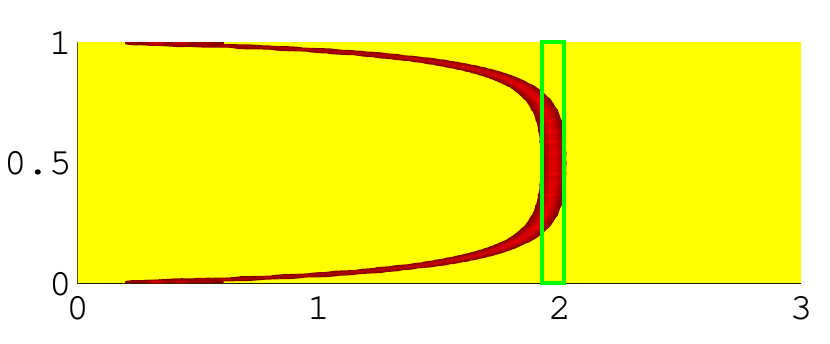}\\
    $(b)~\frac{1}{\mathcal Re}=1,\,\frac{\phi}{\mathcal Re\mathcal Da} = 1$ & $(c)~\frac{1}{\mathcal Re}=10^{-2},\,\frac{\phi}{\mathcal Re\mathcal Da} = 1$\\
    \includegraphics[width=5cm]{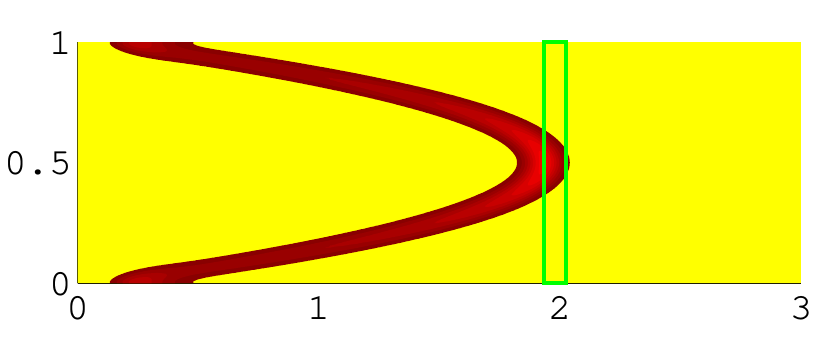}&
    \includegraphics[width=5cm]{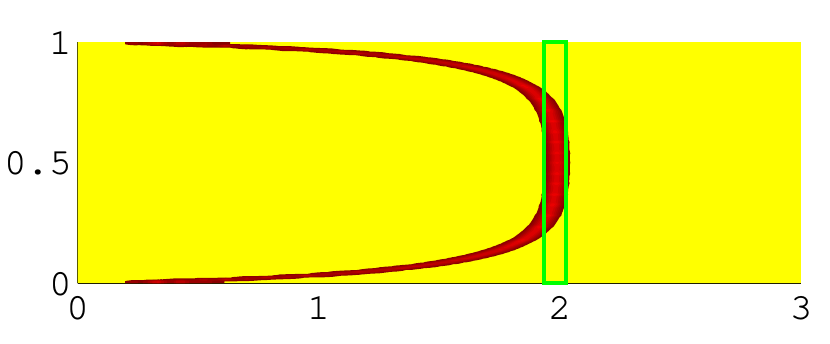}\\
    $(c)~\frac{1}{\mathcal Re}=1,\,\frac{\phi}{\mathcal Re\mathcal Da} = 10^{-6}$ & $(d)~\frac{1}{\mathcal Re}=10^{-2},\,\frac{\phi}{\mathcal Re\mathcal Da} = 10^{-6}$
  \end{tabular}
  \caption{The miscible flow of CO$_2$ and oil has been presented, where red and yellow are used to represent non-zero and zero concentrations of CO$_2$, respectively. $(a)$~The initial concentration, where CO$_2$ has been emplaced in the region of area $A$. $(b)$~The displacement of oil by CO$_2$ at $\mathcal Re =1$, $\mathcal Da = \phi$, and $t=24.4H/U$, where the rectangle represents the region of area $A$, which is expected to be filled with CO$_2$. ($c$)~$\mathcal Re=10^2$, $\mathcal Da = \phi$,  and $t=2.3H/U$. $(d)$~$\mathcal Re = 1$, $\mathcal Da =\phi\times 10^6$, and $t=22.5H/U$. $(e)$~$\mathcal Re = 10^2$, $\mathcal Da =\phi\times 10^6$, and $t=2.28H/$U.}
  \label{fig:nuv1}
\end{figure}

Let us now study the effect of reservoir permeability. Comparing Fig.~\ref{fig:nuv1}$(b$-$c)$ with Fig.~\ref{fig:nuv1}$(d$-$e)$ indicate that reducing the damping force of the porous media, {\em i.e.} the drag, by a factor of $10^6$ has no effect on controlling the shape of the moving sample of CO$_2$. Since an upscaling model is to approximate the detailed flow in pores of the porous matrix, the role of the viscous stress is more likely to resolve the shearing effect in addition to the effect of porous media. However, most studies of miscible flow in a porous media put emphasis on determining how the drag of the porous media depends on $\mu(c)$, ignoring the viscous stress, where $\mu(c)$ appears only in the term that models the drag force. Moreover, to enhance the oil recovery process through a miscible displacement, it is also important to understand the necessary conditions for which a rectangular sample of CO$_2$ would migrate without much distortion of its initial shape. In the Darcy's model~(\ref{eq:dc}), reducing $\mu(c)$ by a factor of $10$ is equivalent to increasing the permeability by the same factor. An increased permeability would enhance the rate of momentum transfer, thereby requiring a balance by the shearing stress.  
Since mixing and dissolution occurs at the molecular label, increasing the rate of momentum transfer would enhance the level of mixing; however, the slowly moving CO$_2$ molecules near a solid body will have more chance to be dissolved. Thus, the treatment of mixing and dissolution in a homogeneous porous media should be different near an impermeable region or in a heterogeneous porous media.%

\begin{figure}[htbp]
  \centering
  \begin{tabular}{cc}
    $\mathcal Re=1$ & $\mathcal Re=10^2$ \\
    \includegraphics[height=2.5cm]
    {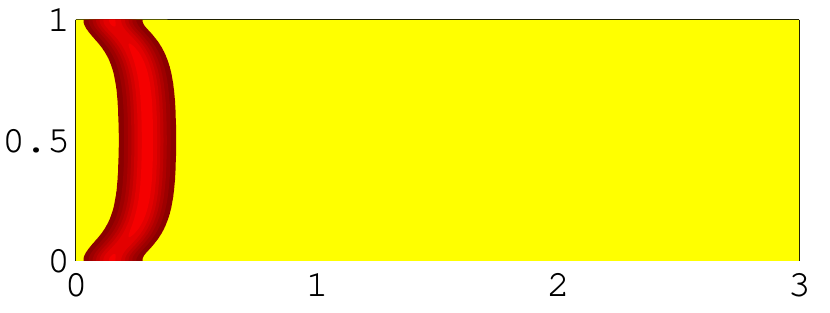}
    &
    \includegraphics[height=2.5cm]
    {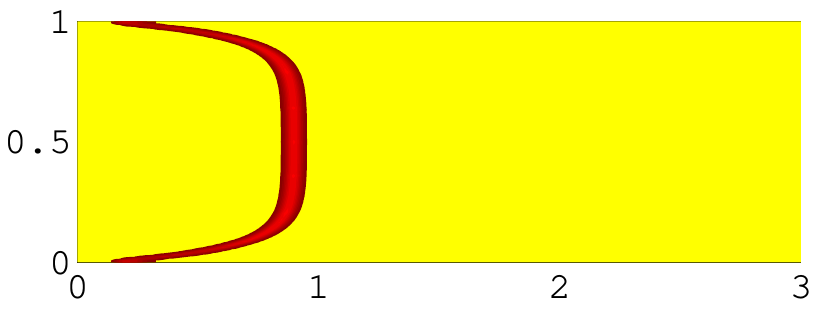}
    \\
    $(a)$ $t=30H/U$, $\mathcal Da=\phi\times 10^{-2}$& $(d)$ $t=2H/U$, $\mathcal Da=\phi\times 10^{-2}$\\
    \includegraphics[height=2.5cm]
    {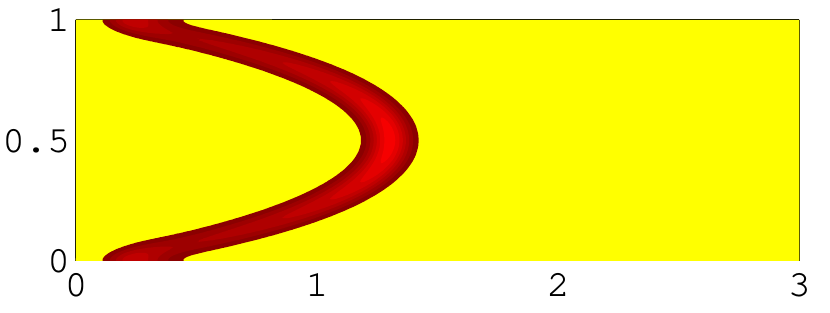}
    &
    \includegraphics[height=2.5cm]
    {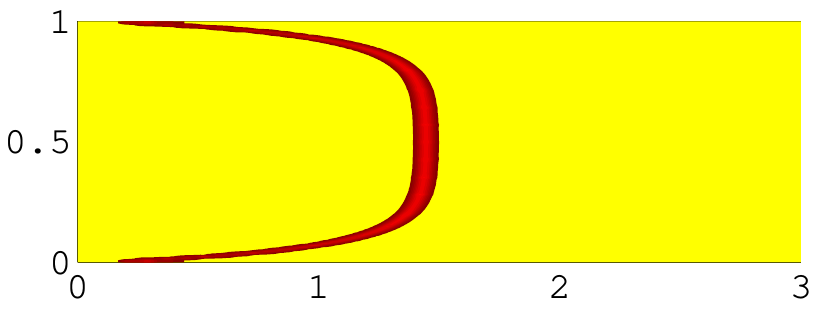}
    \\
    $(b)$ $t=30H/U$, $\mathcal Da=\phi\times 10^{-1}$& $(e)$ $t=2H/U$, $\mathcal Da=\phi\times 10^{-1}$\\

    \includegraphics[height=2.5cm]
    {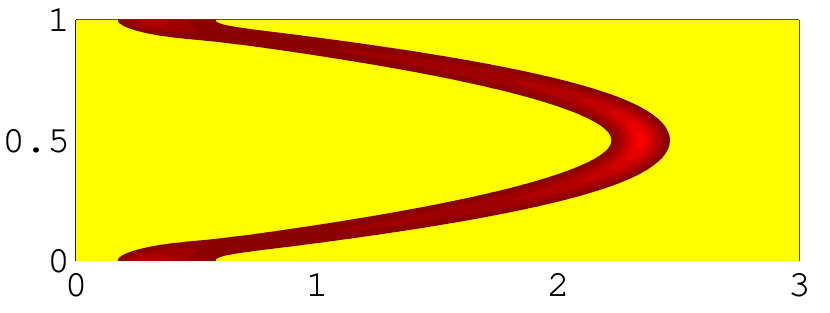}
    &
    \includegraphics[height=2.5cm]
    {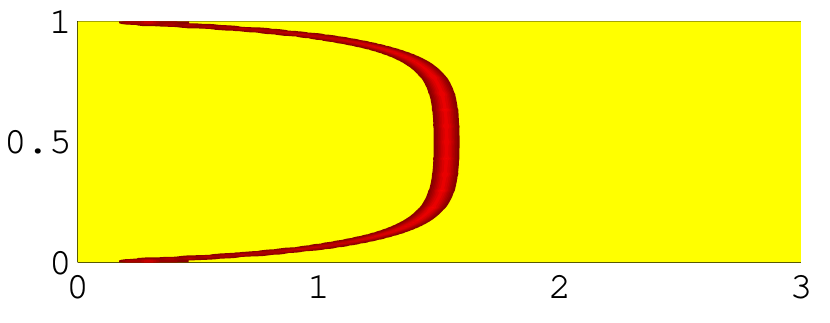}
    \\
    $(c)$ $t=30H/U$, $\mathcal Da=\phi$& $(f)$ $t=2H/U$, $\mathcal Da=\phi$\\
  \end{tabular}
  \caption{Temporal evolution of the flow with respect to variations of $\mathcal Re$ and $\mathcal Da$ with $\mathcal Re\mathcal Sc=20\,000$ and $\phi=18\%$. Left column ($a,\,b,\,c$), $\mathcal Re=1,\,t=30H/U$. Right column ($d,\,e,\,f$), $\mathcal Re=10^2,\,t=2H/U$. }
  \label{fig:da}
\end{figure}

To have further insight into the effect of permeability, two sets of results are displayed in Fig.~\ref{fig:da} for $\mathcal Da/\phi=10^{-2}$, $10^{-1}$, and $1$ at two values of $\mathcal Re$. Plots in Fig.~\ref{fig:da} indicate that CO$_2$ moves faster if the permeability increases - as expected. However, this speed up is also influenced strongly by $\mathcal Re$, and the distortion of the CO$_2$ sample into a parabolic shape is not controlled by the permeability -- the shape of the moving CO$_2$ sample depends on $\mathcal Re$. This indicates that hydrodynamic dispersion is important.

\subsection{Dispersion phenomena in a miscible displacement}
\begin{figure}[htbp]
  \centering
  \begin{tabular}{cc}
    \multicolumn{2}{c}{
      \includegraphics[width=7cm]{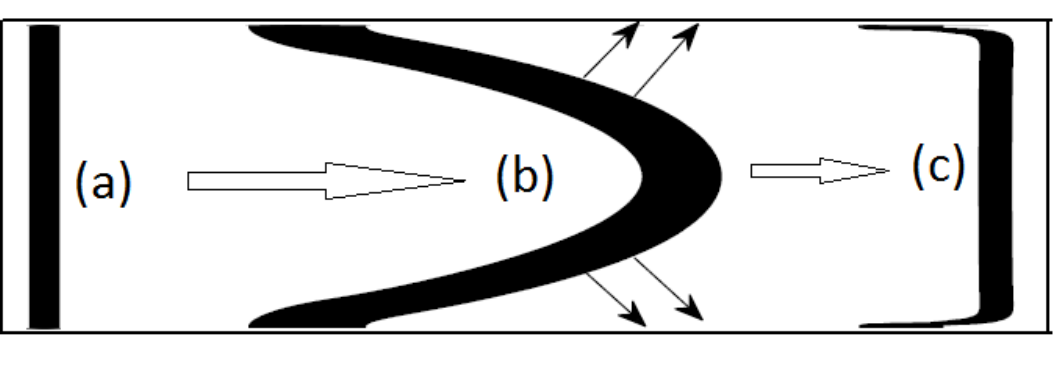}
    }\\
    \multicolumn{2}{c}{($d$) A conceptual model for dispersion}\\
    \includegraphics[height=4.25cm] {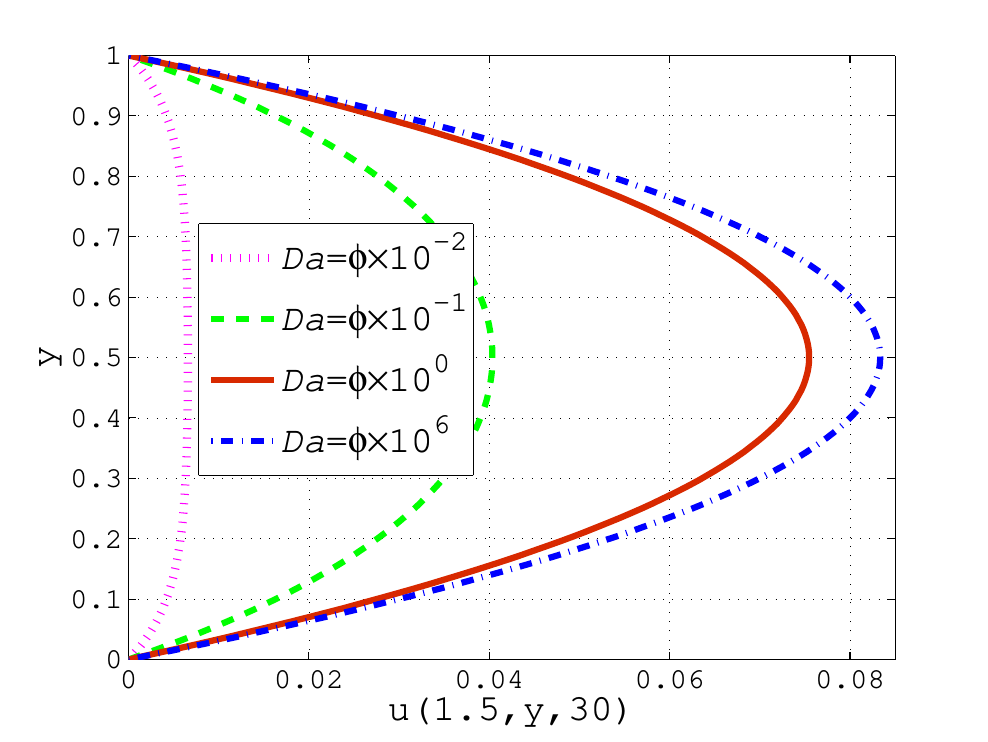}  &
    \includegraphics[height=4.25cm] {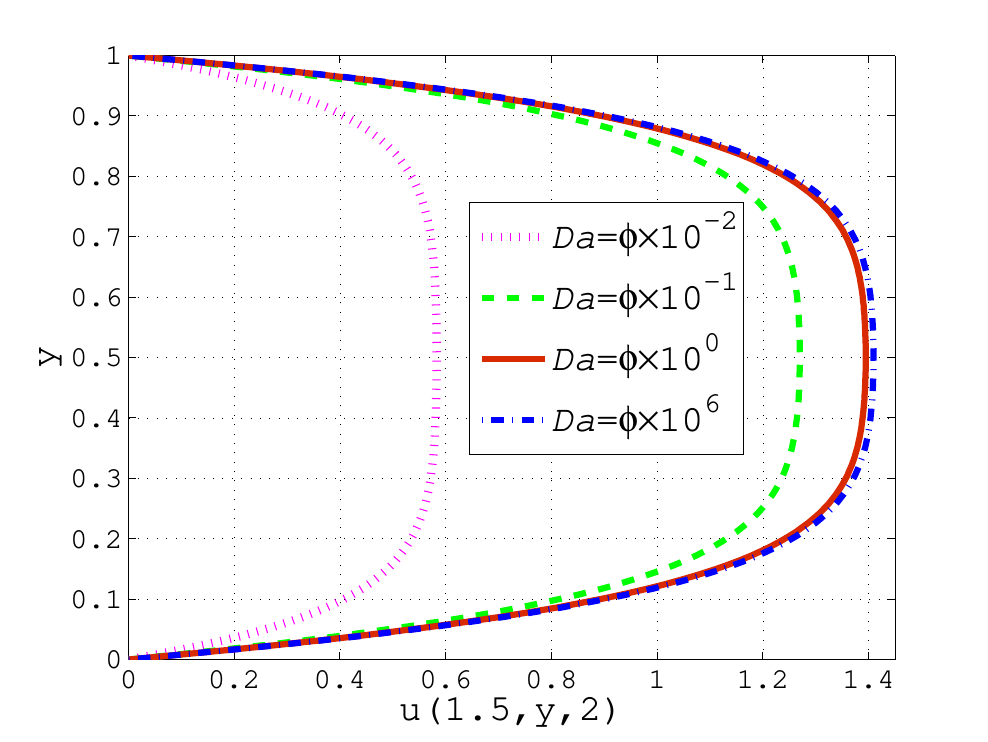}  \\
    $(e)\, u_1(1.5,x_2)$ at $\mathcal Re=1$ & $(f)\, u_1(1.5,x_2)$ at $\mathcal Re=10^2$\\
    \includegraphics[height=4.25cm] {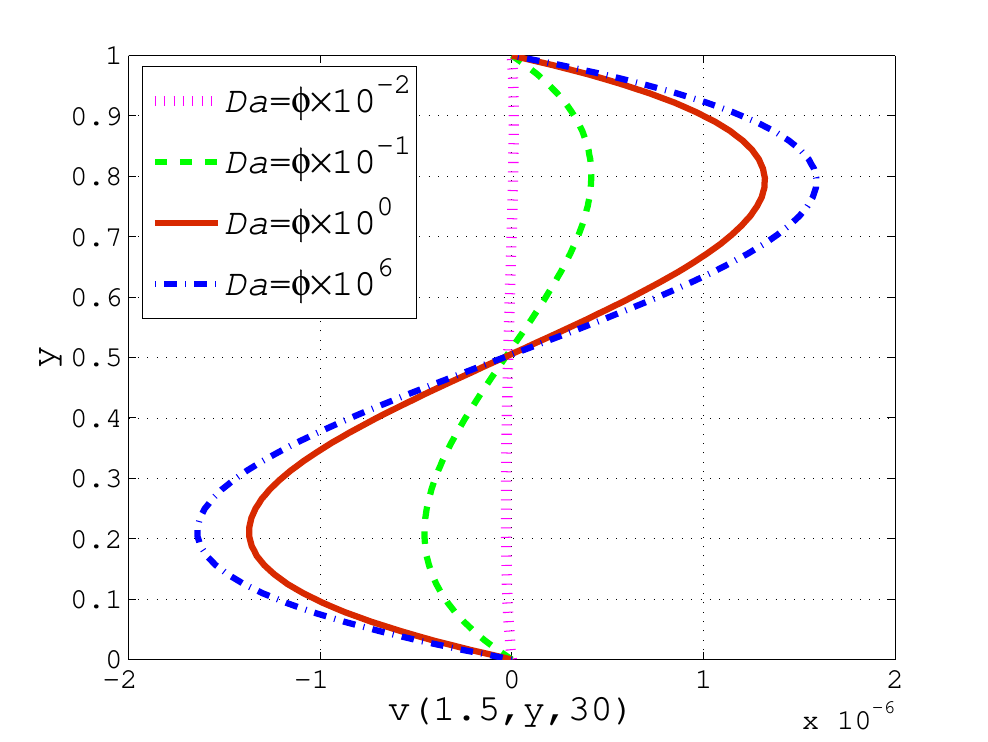}  &
    \includegraphics[height=4.25cm] {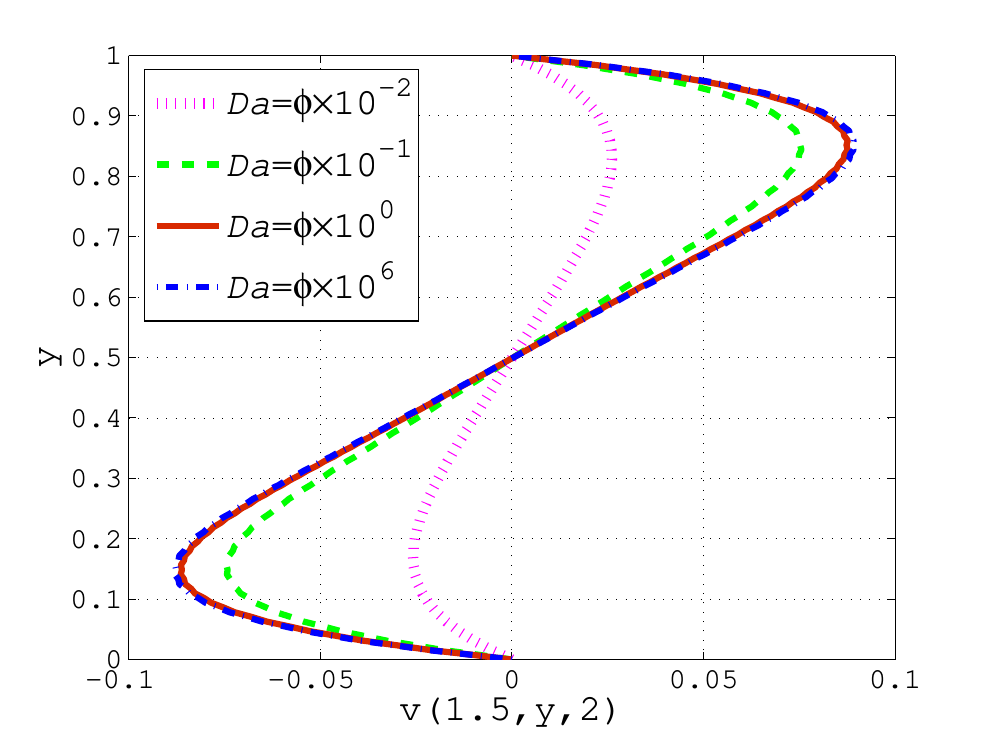}  \\
    $(g)\, u_2(1.5,x_2)$ at $\mathcal Re=1$ & $(h)\, u_2(1.5,x_2)$ at $\mathcal Re=10^2$
  \end{tabular}
  \caption{$(d)$~A schematic illustration for the hydrodynamic dispersion. A flat band of CO$_2$ would take the parabolic shape in the absence transverse velocity as shown by arrows in $(b)$. Velocity profiles $u_1(1.5,x_2)$ and $u_1(1.5,x_2)$ for $\mathcal Re=1$ and $\mathcal Re=10^2$ and $10^{-2}\le\mathcal Da/\phi\le 10^6$.}
  \label{fig:vxda}
\end{figure}

For a fluid flow through an isotropic porous media, the dispersion depends largely on the velocity field \cite{Koch85,Hsu90}. For enhancing the oil recovery with a miscible displacement approach, both the shape and the timing of arrival of CO$_2$ at the production well are of paramount importance. An initially flat band of CO$_2$ ({\em e.g.} Fig.~\ref{fig:nuv1}($a$)) may be dispersed or diffused as it travels in the reservoir. This has been illustrated schematically in Fig.~\ref{fig:vxda}$(d)$, indicating the expected velocity vectors in order for keeping a flat band instead of the parabolic shape. To sweep the oil out of the reservoir, a flat band of CO$_2$ would optimize the oil recovery, in which a positive vertical velocity near the upper boundary and a negative vertical velocity near the bottom boundary would prevent from the parabolic bending of CO$_2$. 

Fig.~\ref{fig:vxda} presents the horizontal and the vertical velocity profiles, $u_i(1.5,y,T)$ ($i=1,\,2$), where $T=30H/U$ for $\mathcal Re=1$ and $T=2H/U$ for $\mathcal Re=10^2$. For $\mathcal Re=1$ the flow reached a steady state when $tH/U=0.8$ if $\mathcal Da/\phi=1$. The smaller the $\mathcal Da$ the earlier the flow reached the steady state. When $\mathcal Re=1$, the inertial effect is balanced by the viscous stress, and the pressure gradient force is balanced by the drag force exerted by the porous media if $\mathcal Da/\phi=1$. The drag force dominates if $\mathcal Da/\phi < 1$ and the pressure gradient dominates if $\mathcal Da/\phi > 1$. We have noticed that the flow remains unsteady if $\mathcal Re=10^2$, where inertial effect is dominant over the viscous stress, and the pressure gradient is balanced by the drag of the porous media if $\mathcal Da/\phi=10^{-2}$.  One also notices from Fig.~\ref{fig:vxda} that the parabolic nature of the horizontal velocity profile turns to a relatively flat shape, when the nonlinear inertial effect in the porous medium dominates, and the strength of the horizontal flow decreases if $\mathcal Da$ decreases. The vertical velocity profiles indicate clearly that the hydrodynamic transverse dispersion is influenced by the dominant inertial effect. Furthermore, the velocity profiles in Fig.~\ref{fig:vxda}($g,h$) show that the displacement of CO$_2$ as a flat band is enhanced by about a factor of $10^5$ if the viscous stress is reduced by a factor of $10^2$.
\subsection{The effect of solvent dissolution}
In this section, results of numerical simulation on the effect of CO$_2$ dissolution have been summarized. Note that in the dimensionless form, the drag force 
associated with the dissolution of the solvent takes the form
$$
f_i^s = \frac{C_{\varphi}}{\lambda_c^2\mathcal Re\mathcal Da}\mu^c\frac{\partial\varphi}{\partial x_i},
$$
where $C_{\varphi}$ is a model parameter.

\begin{figure}
  \centering
  \begin{tabular}{ccc}
    \includegraphics[height=2cm]{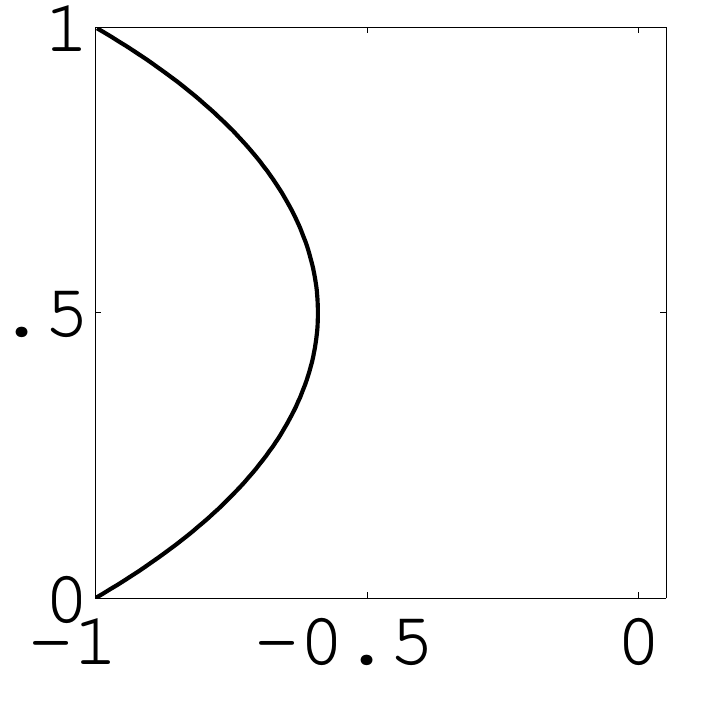}&
    \includegraphics[height=2cm]{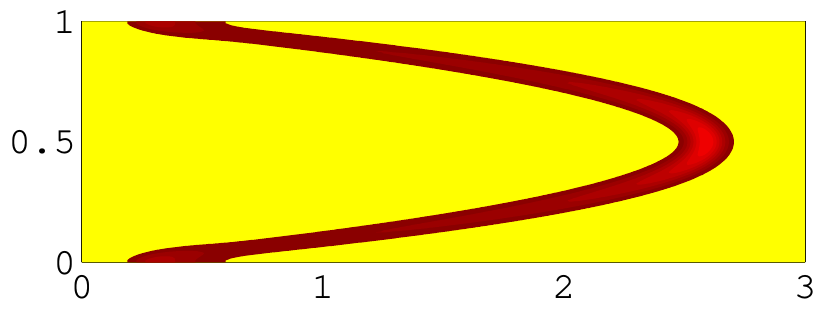}&
    \includegraphics[height=2cm]{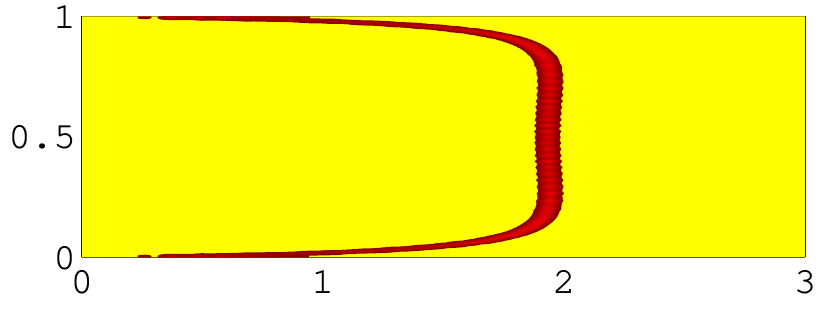}\\
    \includegraphics[height=2cm]{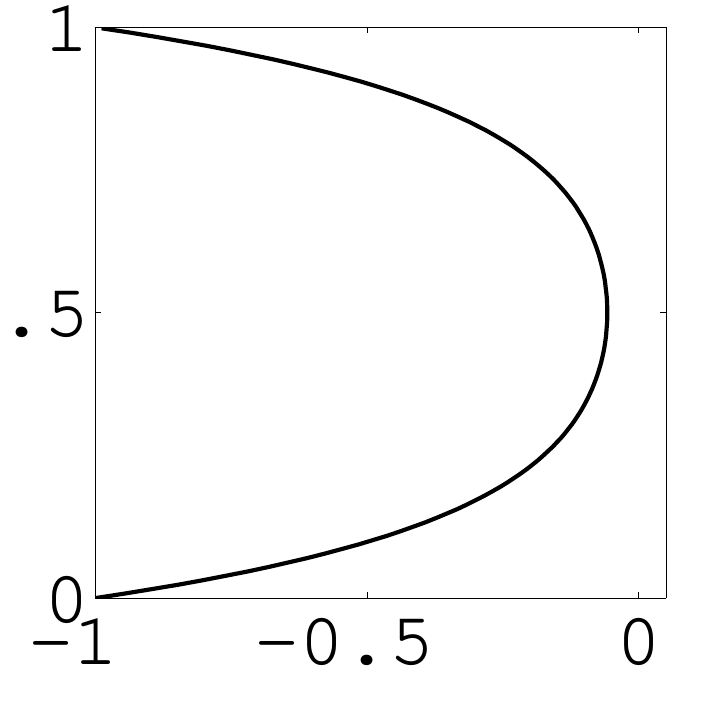}&
    \includegraphics[height=2cm]{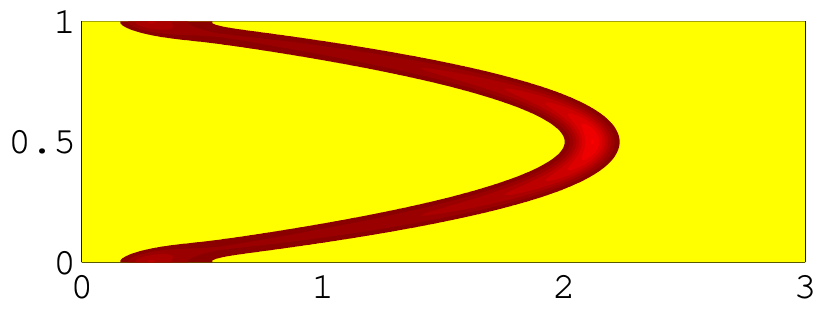}&
    \includegraphics[height=2cm]{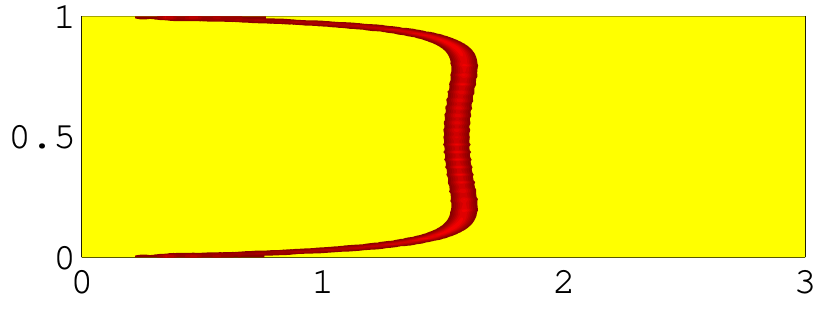}\\
    \includegraphics[height=2cm]{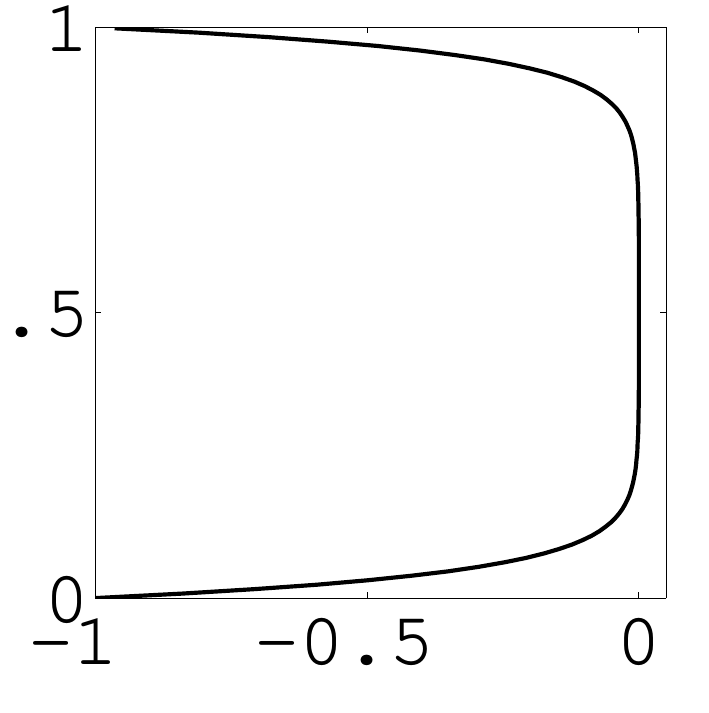}&
    \includegraphics[height=2cm]{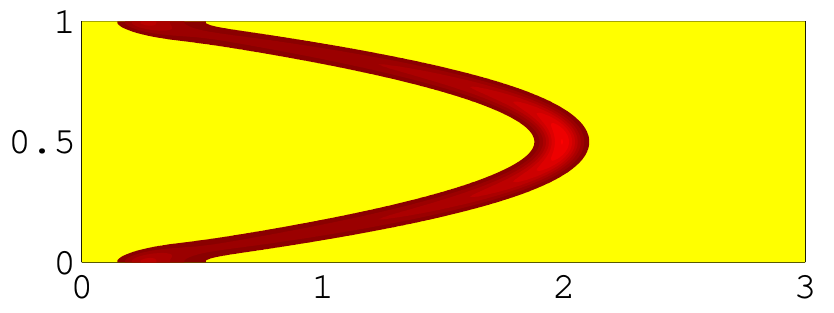}&
    \includegraphics[height=2cm]{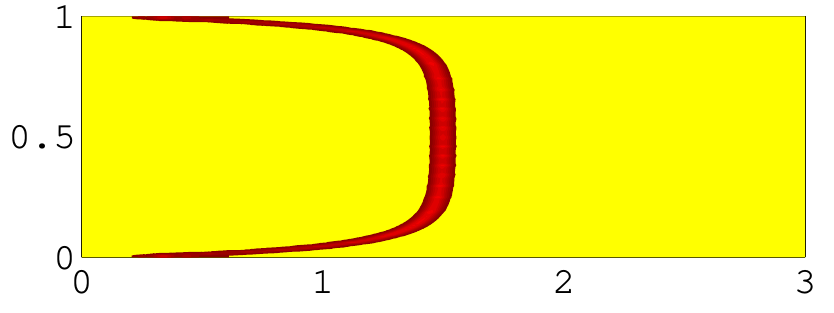}\\
    $(a)$ $\mu^c$ & $(b)~\mathcal Re=1,\,\mathcal Da=1$  & $(c)~\mathcal Re=10^2,\,\mathcal Da=10^{-2}$
  \end{tabular}
  \caption{Results with, top to bottom, $\lambda_c^2=2\times 10^{-1}$, $2\times 10^{-2}$, and $2\times 10^{-3}$. $(a)~\mu^c(1.5,y)$; a thin layer of width~$\sim\lambda_c$ is seen adjacent to the top and bottom boundaries. $(b)~c(x,y,25)$ and $(c)~c(x,y,3)$.}
  \label{fig:A1}
\end{figure}

The first question is to investigate  the role of $\mu^c$ when $\lambda_c$ decreases but the overall strength of $f_i^s$ remains the same for each value of $\lambda_c$. Thus, we have carried out numerical experiments by decreasing the value of $\lambda_c$ with $\mathcal Re=1$, $\mathcal Da=1$, and  $C_{\varphi}=\lambda_c^2$, and then repeated these experiments with various other values of $\mathcal Re$ and $\mathcal Da$ for each value of $\lambda_c$. For each case, $\frac{C_{\varphi}}{\lambda_c^2\mathcal Re\mathcal Da}=1$ has been used. Results with $\lambda_c^2=2\times 10^{-1}$, $2\times 10^{-2}$, and $2\times 10^{-3}$ are presented in Fig.~\ref{fig:A1}, where the first column~(Fig.~\ref{fig:A1}$a$) shows $\mu^c(1.5,y)$. Clearly, the influence of $\mu^c$ would accelerate the flow near each of the impermeable boundaries in a region of width $\lambda_c$. Having $C_{\varphi}/(\lambda_c^2\mathcal Re\mathcal Da) = 1$, we are able to examine how a spatially varying viscosity $\mu^c$ influences mass and momentum transfer in a porous media. The plots of $c(x_1,x_2,25)$ and $c(x_1,x_2,3)$ in Figs~\ref{fig:A1}$b$ and~\ref{fig:A1}$c$, respectively, show a speed up in arriving CO$_2$ toward the production well if $\lambda_c$ increases.  However, for a small $\lambda_c$, we see that $\mu^c$ is nearly zero except in a narrow region that is adjacent to the boundaries. 

\begin{figure}
  \centering
  \begin{tabular}{l}
    \includegraphics[width=5.2cm]{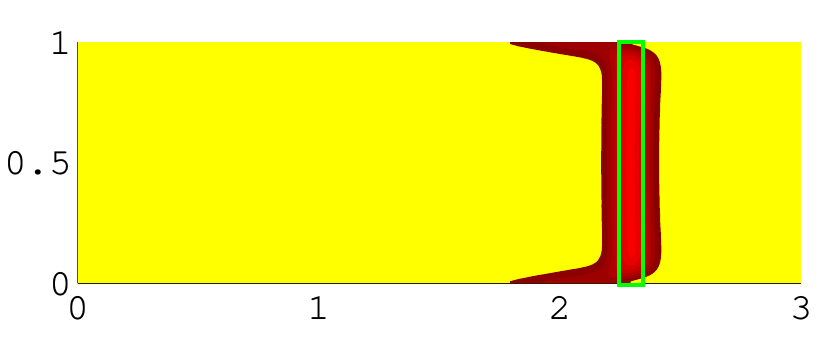}\\
    $\quad(a)~\mathcal Sc=10^4$\\
    \includegraphics[height=2.2cm]{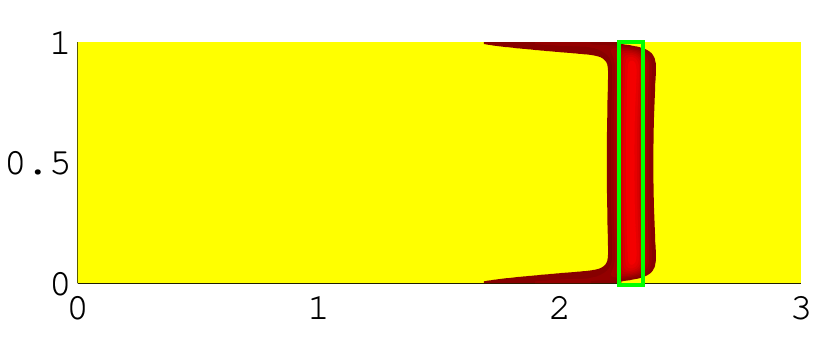}\\
    $\quad(b)~\mathcal Sc=2\times 10^4$\\
    \includegraphics[height=2.2cm]{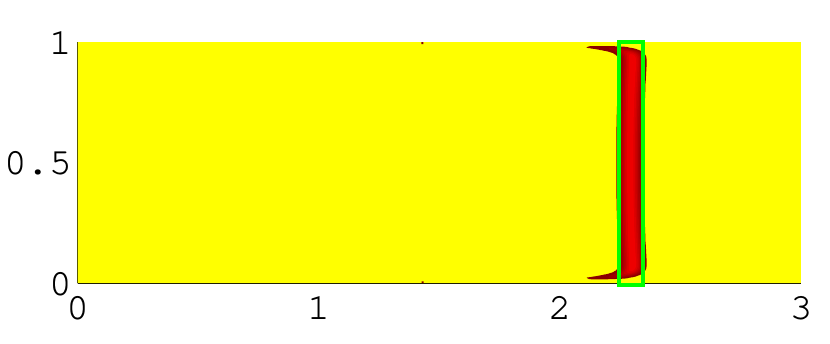}\\
    $\quad(c)~\mathcal Sc=10^5$\\
    \includegraphics[trim=3.0cm 7.5cm 3.5cm 8cm,clip=true,height=4.5cm,width=5.25cm]{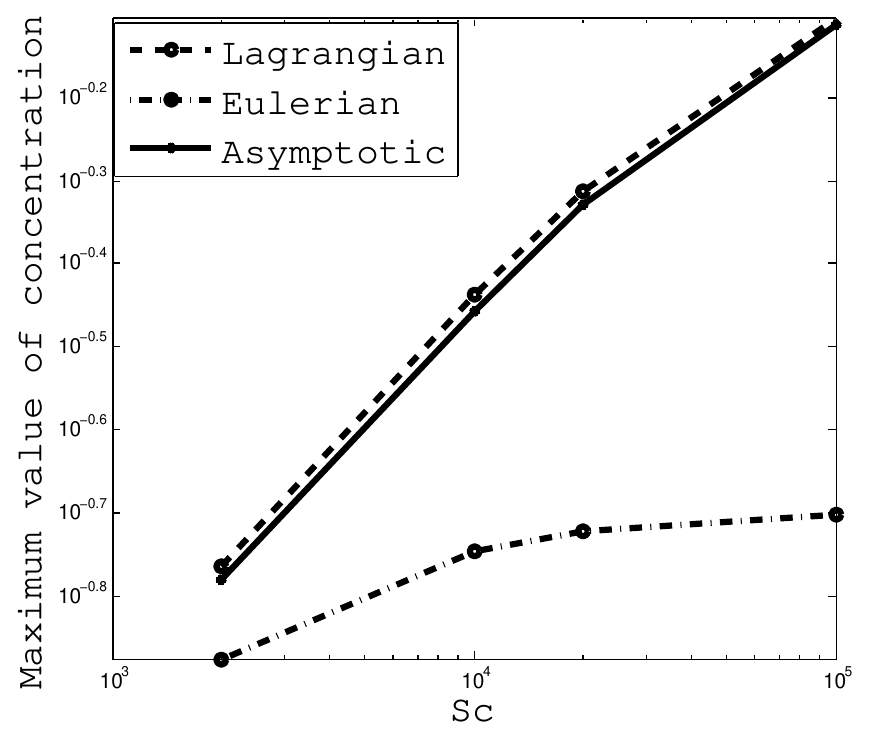}
  \end{tabular}
  \caption{$(a$-$c)$ The migration of CO$_2$ as a flat band at $\mathcal Sc=10^4,\,2\times 10^4,\, 10^5$, when the viscosity is reduced near the reservoir boundaries. $(d)$ Maximum value of the concentration has a good agreement with asymptotic results. However, the reference Eulerian simulation does not converge to the present Lagrangian simulation.}
  \label{fig:pstn}
\end{figure}

To investigate the potential of the present statistical mechanical approach, results with simulations at $\lambda_c^2=7.1\times 10^{-5}$, $\mathcal Sc=10^4$, $2\times 10^4$, $10^5$, and $C_{\varphi}=0.42$ have been presented in Fig.~\ref{fig:pstn}. Note that the larger the $\mathcal Sc$ the smaller the hydrodynamic dispersion $\mathcal D (\unit{m^2}{\cdot}\unit{s^{-1}})$ of CO$_2$ in oil. The value of $\mathcal D$ may be determined empirically, based on field measurements ({\em e.g.}~\cite{Gelhar92}), and clearly, the present development could be used to validate such empirical values. Fig.~\ref{fig:pstn}$(c)$ shows that for $\mathcal D=3.5\times 10^{-8}~\unit{m^2}{\cdot}\unit{s^{-1}}$ the dispersion of CO$_2$ is negligible until $t=25H/U$, which is indicated by the rectangle in Fig.~\ref{fig:pstn}. 

One important benefit of using the Lagrangian approach for the present development has been explored in Fig.~\ref{fig:pstn}($d$). Here, we have compared the maximum values of the concentration for each plot in Fig.~\ref{fig:pstn}$(a$-$c)$ with an approximate solution of~(\ref{eq:mnc}), which is derived from asymptotic analysis. This data has also been compared with a reference model, where eq.(\ref{eq:mnc}) has been solved with a complete Eulerian scheme. It is clear from Fig.~\ref{fig:pstn}($d$) that the Eulerian solution does not converge to the asymptotic solution. 

\section{Conclusion and future direction}\label{sec:dis}

There have been great interests on miscible displacement of oil by a solvent in the field of petroleum engineering, and the topic of CO$_2$ enhanced oil recovery has received an increased attention. Recently, there is a growing propensity on employing high performance CFD reservoir simulation techniques with the hope that this may discover substantial currently unrecognized opportunities for increasing the economic recovery of hydrocarbons\cite{Jenny2006}. A detailed CFD investigation helps understand how injected CO$_2$ displaces oil in the pore space of rocks, and provides further insight when field results fall short of laboratory performance. However, the field scale sweep efficiency is one of the most important factors affecting the economic recovery of hydrocarbons, for which an improved upscaling model may be beneficial to industries instead of studying the details of pore scale flow. This article reiterates such a growing trend, and outlines some significant development in this direction.

To analyze the sweep efficiency for a miscible displacement, an overall flow characteristics can be studied through an upscaling model. This article has focused on the development of a generalized upscaling model employing a statistical mechanical approach to resolve the effect of CO$_2$ dissolution, and studied factors for optimizing the pressure drag and skin friction which are exerted by the porous media at the reservoir scale. In particular, the idealized influence of isolated impermeable region partially embedded in a porous matrix has been studied. Note, the averaging process employed in this article is commonly used to model turbulent flow in a porous media. We have outlined how to extend and re-design an intrinsic space-time average operator to model small scale transient variability in a laminar miscible flow, which is a distinct fundamental development of this work. This research has clearly explained the regime of the viscous shearing stress, which play a negligible role on large scale reservoir flow; however, is important for accurate upscaling of small scale physics. 

Some important future developments include the following. An extension to a 3D multilevel mesh employing a massively parallel algorithm would add potential benefits toward needs of petroleum industries. We are currently developing data structures using the object oriented C++ programming paradigm. The present results clearly put hints for using adaptive mesh approaches~\cite{Wirasaet2005,Alam2006}. However, a full understanding of the efficiency of multilevel and Lagrangian approaches are essential before studying the efficiency of parallel adaptive mesh methods for such a miscible flow problem.

\section*{Acknowledgments}
JMA acknowledges financial support from the National Science and Research Council~(NSERC), Canada.

%% file: alamjporous2014.bbl
\begin{thebibliography}{56}
\expandafter\ifx\csname natexlab\endcsname\relax\def\natexlab#1{#1}\fi
\providecommand{\bibinfo}[2]{#2}
\ifx\xfnm\relax \def\xfnm[#1]{\unskip,\space#1}\fi
\bibitem[{Zhao et~al.(2012)Zhao, Cheng, and Zhang}]{Zhao2012}
\bibinfo{author}{R.~Zhao}, \bibinfo{author}{J.~Cheng},
  \bibinfo{author}{K.~Zhang},
\newblock \bibinfo{title}{{CO$_2$} plume evolution and pressure buildup of
  large-scale {CO}$_2$ injection into saline aquifers in sanzhao depression,
  songliao basin, china},
\newblock \bibinfo{journal}{Transport in Porous Media} \bibinfo{volume}{95}
  (\bibinfo{year}{2012}) \bibinfo{pages}{407--424}.
\bibitem[{Leetaru et~al.(2009)Leetaru, Frailey, Damico, Mehnert, Birkholzer,
  Zhou, and Jordan}]{Leetaru2009}
\bibinfo{author}{H.~E. Leetaru}, \bibinfo{author}{S.~M. Frailey},
  \bibinfo{author}{J.~Damico}, \bibinfo{author}{E.~Mehnert},
  \bibinfo{author}{J.~Birkholzer}, \bibinfo{author}{Q.~Zhou},
  \bibinfo{author}{P.~D. Jordan},
\newblock \bibinfo{title}{Understanding {CO$_2$} plume behavior and basin-scale
  pressure changes during sequestration projects through the use of reservoir
  fluid modeling},
\newblock \bibinfo{journal}{Energy Procedia} \bibinfo{volume}{1}
  (\bibinfo{year}{2009}) \bibinfo{pages}{1799 -- 1806}.
\bibitem[{Mohsin and Anazi(2009)}]{Mohsin2009}
\bibinfo{author}{A.~J. Mohsin}, \bibinfo{author}{A.~B.~D. Anazi},
\newblock \bibinfo{title}{A comparison study of the of the {CO}$_2$-oil
  physical properties-literature correlations accuracy using visual basic
  modeling technique},
\newblock \bibinfo{journal}{Oil and Gas Business} \bibinfo{volume}{60(5)}
  (\bibinfo{year}{2009}) \bibinfo{pages}{287--291}.
\bibitem[{Nobakht et~al.(2007)Nobakht, Moghadam, and Gu}]{Nobakht2007}
\bibinfo{author}{M.~Nobakht}, \bibinfo{author}{S.~Moghadam},
  \bibinfo{author}{Y.~Gu},
\newblock \bibinfo{title}{Effects of viscous and capillary forces on {CO}$_2$
  enhanced oil recovery under reservoir conditions},
\newblock \bibinfo{journal}{Energy \& Fuels} \bibinfo{volume}{21}
  (\bibinfo{year}{2007}) \bibinfo{pages}{3469 -- 3476}.
\bibitem[{Sen(2008)}]{Sen2008}
\bibinfo{author}{R.~Sen},
\newblock \bibinfo{title}{Biotechnology in petroleum recovery: The microbial
  {EOR}},
\newblock \bibinfo{journal}{Progress in Energy and Combustion Science}
  \bibinfo{volume}{34} (\bibinfo{year}{2008}) \bibinfo{pages}{714 -- 724}.
\bibitem[{Peaceman and Rachford(1962)}]{Peaceman62}
\bibinfo{author}{D.~W. Peaceman}, \bibinfo{author}{H.~H. Rachford},
\newblock \bibinfo{title}{Numerical calculation of the multidimensional
  miscible displacement},
\newblock \bibinfo{journal}{Society of Petroleum Engineering Journal}
  \bibinfo{volume}{2} (\bibinfo{year}{1962}) \bibinfo{pages}{327--339}.
\bibitem[{Blackwell et~al.(1959)Blackwell, Rayne, and Terry}]{Blackwell59}
\bibinfo{author}{R.~R. Blackwell}, \bibinfo{author}{J.~R. Rayne},
  \bibinfo{author}{W.~M. Terry},
\newblock \bibinfo{title}{Fectors influencing the efficiency of miscible
  displacement},
\newblock \bibinfo{journal}{{AIME} {P}etroleum {T}ransactions}
  \bibinfo{volume}{217} (\bibinfo{year}{1959}).
\bibitem[{Chen and Wilkinson(1985)}]{Chen85}
\bibinfo{author}{J.-D. Chen}, \bibinfo{author}{D.~Wilkinson},
\newblock \bibinfo{title}{Pore-scale viscous fingering in porous media},
\newblock \bibinfo{journal}{Phys. Rev. Lett.} \bibinfo{volume}{55}
  (\bibinfo{year}{1985}) \bibinfo{pages}{1892--1895}.
\bibitem[{Homsy(1987)}]{Homsy87}
\bibinfo{author}{G.~M. Homsy},
\newblock \bibinfo{title}{Viscous fingering in porous media},
\newblock \bibinfo{journal}{Annual Review of Fluid Mechanics}
  \bibinfo{volume}{19} (\bibinfo{year}{1987}) \bibinfo{pages}{271--311}.
\bibitem[{Tan and Homsy(1988)}]{Tan88}
\bibinfo{author}{C.~T. Tan}, \bibinfo{author}{G.~M. Homsy},
\newblock \bibinfo{title}{Simulation of nonlinear viscous fingering in miscible
  displacement},
\newblock \bibinfo{journal}{Physics of Fluids} \bibinfo{volume}{31}
  (\bibinfo{year}{1988}) \bibinfo{pages}{1330--1338}.
\bibitem[{Lenormand(1989)}]{Lenormand89}
\bibinfo{author}{R.~Lenormand},
\newblock \bibinfo{title}{Flow through porous media: Limits of fractal
  patterns},
\newblock \bibinfo{journal}{Proceedings of the Royal Society of London. Series
  A, Mathematical and Physical Sciences} \bibinfo{volume}{423}
  (\bibinfo{year}{1989}) \bibinfo{pages}{pp. 159--168}.
\bibitem[{Jha et~al.(2011)Jha, Cueto-Felgueroso, and Juanes}]{Birendra2011}
\bibinfo{author}{B.~Jha}, \bibinfo{author}{L.~Cueto-Felgueroso},
  \bibinfo{author}{R.~Juanes},
\newblock \bibinfo{title}{Fluid mixing from viscous fingering},
\newblock \bibinfo{journal}{Phys. Rev. Lett.} \bibinfo{volume}{106}
  (\bibinfo{year}{2011}) \bibinfo{pages}{194502}.
\bibitem[{Longmuir(2004)}]{Gavin2004}
\bibinfo{author}{G.~Longmuir},
\newblock \bibinfo{title}{Pre-darcy flow: a missing piece of the improved oil
  recovery puzzle?},
\newblock \bibinfo{journal}{Society of Petroleum Engineers}
  (\bibinfo{year}{2004}) \bibinfo{pages}{SPE 89433}.
\bibitem[{Yu et~al.(2012)Yu, Bian, Li, Zhang, Yan, Wang, and Wang}]{Rongze2012}
\bibinfo{author}{R.~Yu}, \bibinfo{author}{Y.~Bian}, \bibinfo{author}{Y.~Li},
  \bibinfo{author}{X.~Zhang}, \bibinfo{author}{J.~Yan},
  \bibinfo{author}{H.~Wang}, \bibinfo{author}{K.~Wang},
\newblock \bibinfo{title}{Non-darcy flow numerical simulation of xpj low
  permeability reservoir},
\newblock \bibinfo{journal}{Journal of Petroleum Science and Engineering}
  \bibinfo{volume}{92-93} (\bibinfo{year}{2012}) \bibinfo{pages}{40--47}.
\bibitem[{Garibotti and Peszyska(2009)}]{Garibotti2009}
\bibinfo{author}{C.~Garibotti}, \bibinfo{author}{M.~Peszyska},
\newblock \bibinfo{title}{Upscaling non-darcy flow},
\newblock \bibinfo{journal}{Transport in Porous Media} \bibinfo{volume}{80}
  (\bibinfo{year}{2009}) \bibinfo{pages}{401--430}.
\bibitem[{Bear(1972)}]{Bear72}
\bibinfo{author}{J.~Bear}, \bibinfo{title}{Dynamics of fluids in porous media},
  \bibinfo{publisher}{Elsevier (New York)}, \bibinfo{year}{1972}.
\bibitem[{Zeng and Grigg(2006)}]{Zeng2006}
\bibinfo{author}{Z.~Zeng}, \bibinfo{author}{R.~Grigg},
\newblock \bibinfo{title}{A criterion for non-darcy flow in porous media},
\newblock \bibinfo{journal}{Transport in Porous Media} \bibinfo{volume}{63}
  (\bibinfo{year}{2006}) \bibinfo{pages}{57--69}.
\bibitem[{Fancher and Lewis(1933)}]{Fancher33}
\bibinfo{author}{G.~H. Fancher}, \bibinfo{author}{J.~A. Lewis},
\newblock \bibinfo{title}{Flow of simple fluids through porous materials},
\newblock \bibinfo{journal}{Ind. Engng. Chem.} \bibinfo{volume}{25(10)}
  (\bibinfo{year}{1933}) \bibinfo{pages}{1139--1147}.
\bibitem[{Green and Duwez(1951)}]{Green51}
\bibinfo{author}{L.~J. Green}, \bibinfo{author}{P.~Duwez},
\newblock \bibinfo{title}{Fluid flow through porous metals},
\newblock \bibinfo{journal}{J. Appl. Mech.}  (\bibinfo{year}{1951})
  \bibinfo{pages}{39--45}.
\bibitem[{Ergun(1952)}]{Ergun52}
\bibinfo{author}{S.~Ergun},
\newblock \bibinfo{title}{Fluid flow through packed columns},
\newblock \bibinfo{journal}{Chem. Engng. Prog} \bibinfo{volume}{48(2)}
  (\bibinfo{year}{1952}) \bibinfo{pages}{89--94}.
\bibitem[{E(1974)}]{Scheidegger74}
\bibinfo{author}{S.~A. E}, \bibinfo{title}{The Physics of Flow through Porous
  Media}, \bibinfo{publisher}{University of Toronto Press},
  \bibinfo{edition}{third} edition, \bibinfo{year}{1974}.
\bibitem[{Blick and Civan(1988)}]{Blick88}
\bibinfo{author}{E.~F. Blick}, \bibinfo{author}{F.~Civan},
\newblock \bibinfo{title}{Porous media momentum equation for highly accelerated
  flow},
\newblock \bibinfo{journal}{SPE Reserv. Engng.}  (\bibinfo{year}{1988})
  \bibinfo{pages}{1048--1052}.
\bibitem[{Meyer and Smith(1985)}]{Meyer85}
\bibinfo{author}{B.~A. Meyer}, \bibinfo{author}{D.~W. Smith},
\newblock \bibinfo{title}{Flow through porous media: comparison of consolidated
  and unconsolidated materials},
\newblock \bibinfo{journal}{Industrial \& Engineering Chemistry Fundamentals}
  \bibinfo{volume}{24} (\bibinfo{year}{1985}) \bibinfo{pages}{360--368}.
\bibitem[{Fand et~al.(1987)Fand, Kim, Lam, and Phan}]{Fand87}
\bibinfo{author}{R.~M. Fand}, \bibinfo{author}{B.~Y.~K. Kim},
  \bibinfo{author}{A.~C.~C. Lam}, \bibinfo{author}{R.~T. Phan},
\newblock \bibinfo{title}{Resistance to the flow of fluids through simple and
  complex porous media whose matrices are composed of randomly packed spheres},
\newblock \bibinfo{journal}{Journal of Fluids Engineering}
  \bibinfo{volume}{109} (\bibinfo{year}{1987}) \bibinfo{pages}{268--273}.
\bibitem[{Hubbert(1957)}]{Hubbert57}
\bibinfo{author}{M.~K. Hubbert},
\newblock \bibinfo{title}{Darcy's law and the field equations of the flow of
  underground fluids},
\newblock \bibinfo{journal}{International Association of Scientific Hydrology.
  Bulletin} \bibinfo{volume}{2} (\bibinfo{year}{1957}) \bibinfo{pages}{23--59}.
\bibitem[{Srinivasan et~al.(2013)Srinivasan, Bonito, and
  Rajagopal}]{Srinivasan2013}
\bibinfo{author}{S.~Srinivasan}, \bibinfo{author}{A.~Bonito},
  \bibinfo{author}{K.~Rajagopal},
\newblock \bibinfo{title}{Flow of a fluid through a porous solid due to high
  pressure gradients},
\newblock \bibinfo{journal}{Journal of Porous Media} \bibinfo{volume}{16}
  (\bibinfo{year}{2013}) \bibinfo{pages}{193--203}.
\bibitem[{Jenny et~al.(2006)Jenny, Lee, and Tchelepi}]{Jenny2006}
\bibinfo{author}{P.~Jenny}, \bibinfo{author}{S.~Lee},
  \bibinfo{author}{H.~Tchelepi},
\newblock \bibinfo{title}{Adaptive fully implicit multi-scale finite-volume
  method for multi-phase flow and transport in heterogeneous porous media},
\newblock \bibinfo{journal}{Journal of Computational Physics}
  \bibinfo{volume}{217} (\bibinfo{year}{2006}) \bibinfo{pages}{627 -- 641}.
\bibitem[{Brinkman(1949)}]{Brinkman49}
\bibinfo{author}{H.~Brinkman},
\newblock \bibinfo{title}{A calculation of the viscous force exerted by a
  flowing fluid on a dense swarm of particles},
\newblock \bibinfo{journal}{Applied Scientific Research} \bibinfo{volume}{1}
  (\bibinfo{year}{1949}) \bibinfo{pages}{27--34}.
\bibitem[{Ma and Ruth(1993)}]{Ma93}
\bibinfo{author}{H.~Ma}, \bibinfo{author}{D.~W. Ruth},
\newblock \bibinfo{title}{The microscopic analysis of high forchheimer number
  flow in porous media},
\newblock \bibinfo{journal}{Transport Porous Media} \bibinfo{volume}{13}
  (\bibinfo{year}{1993}) \bibinfo{pages}{139--160}.
\bibitem[{Whitaker(1996)}]{Whitaker96}
\bibinfo{author}{S.~Whitaker},
\newblock \bibinfo{title}{The forchheimer equation: A theoretical development},
\newblock \bibinfo{journal}{Transport in Porous Media} \bibinfo{volume}{25}
  (\bibinfo{year}{1996}) \bibinfo{pages}{27--61}.
\bibitem[{Koval(1963)}]{Koval63}
\bibinfo{author}{E.~J. Koval},
\newblock \bibinfo{title}{A method for predicting the performance of unstable
  miscible displacement in heterogenous media},
\newblock \bibinfo{journal}{{S}oceity of {P}etroleum {E}ngineers {J}ournal}
  \bibinfo{volume}{450} (\bibinfo{year}{1963}) \bibinfo{pages}{145--154}.
\bibitem[{{Sahu} et~al.(2009){Sahu}, {Ding}, {Valluri}, and {Matar}}]{Sahu2009}
\bibinfo{author}{K.~C. {Sahu}}, \bibinfo{author}{H.~{Ding}},
  \bibinfo{author}{P.~{Valluri}}, \bibinfo{author}{O.~K. {Matar}},
\newblock \bibinfo{title}{{Pressure-driven miscible two-fluid channel flow with
  density gradients}},
\newblock \bibinfo{journal}{Physics of Fluids} \bibinfo{volume}{21}
  (\bibinfo{year}{2009}) \bibinfo{pages}{043603}.
\bibitem[{Jha et~al.(2013)Jha, Cueto-Felgueroso, and Juanes}]{Birendra2013}
\bibinfo{author}{B.~Jha}, \bibinfo{author}{L.~Cueto-Felgueroso},
  \bibinfo{author}{R.~Juanes},
\newblock \bibinfo{title}{Synergetic fluid mixing from viscous fingering and
  alternating injection},
\newblock \bibinfo{journal}{Phys. Rev. Lett.} \bibinfo{volume}{111}
  (\bibinfo{year}{2013}) \bibinfo{pages}{144501}.
\bibitem[{De~Lemos(2006)}]{DeLemos2006}
\bibinfo{author}{M.~De~Lemos}, \bibinfo{title}{Turbulence in Porous Media:
  Modeling And Applications}, \bibinfo{year}{2006}.
\bibitem[{Whitaker(1999)}]{Whitaker99}
\bibinfo{author}{S.~Whitaker}, \bibinfo{title}{The Method of Volume Averaging},
  \bibinfo{publisher}{Springer}, \bibinfo{year}{1999}.
\bibitem[{Lage et~al.(2002)Lage, De~Lemos, and Nield}]{Lage2002}
\bibinfo{author}{J.~Lage}, \bibinfo{author}{M.~De~Lemos},
  \bibinfo{author}{D.~Nield},
\newblock \bibinfo{title}{Modeling turbulence in porous media},
\newblock \bibinfo{journal}{Transport phenomena in porous media}
  \bibinfo{volume}{2} (\bibinfo{year}{2002}) \bibinfo{pages}{198--230}.
\bibitem[{Viswanath et~al.(2007)Viswanath, Ghosh, Prasad, Dutt, and
  Rani}]{Dabir2007}
\bibinfo{author}{D.~S. Viswanath}, \bibinfo{author}{T.~K. Ghosh},
  \bibinfo{author}{D.~H.~L. Prasad}, \bibinfo{author}{N.~V. Dutt},
  \bibinfo{author}{K.~Y. Rani}, \bibinfo{title}{Viscosity of Liquids: Theory,
  Estimation, Experiment, and Data}, \bibinfo{publisher}{Springer},
  \bibinfo{year}{2007}.
\bibitem[{Barus(1893)}]{Barus}
\bibinfo{author}{C.~Barus},
\newblock \bibinfo{title}{Isothermals, isopiestics and isometrics relative to
  viscosity},
\newblock \bibinfo{journal}{Americal Journal of Science} \bibinfo{volume}{XLV}
  (\bibinfo{year}{1893}).
\bibitem[{Powell et~al.(1941)Powell, Roseveare, and Eyring}]{Eyring41}
\bibinfo{author}{R.~E. Powell}, \bibinfo{author}{W.~E. Roseveare},
  \bibinfo{author}{H.~Eyring},
\newblock \bibinfo{title}{Diffusion, thermal conductivity, and viscous flow of
  liquids},
\newblock \bibinfo{journal}{Industrial \& Engineering Chemistry}
  \bibinfo{volume}{33} (\bibinfo{year}{1941}) \bibinfo{pages}{430--435}.
\bibitem[{Eyring(1936)}]{Eyring36}
\bibinfo{author}{H.~Eyring},
\newblock \bibinfo{title}{Viscosity, plasticity, and diffusion as examples of
  absolute reaction rates},
\newblock \bibinfo{journal}{The Journal of Chemical Physics}
  \bibinfo{volume}{4} (\bibinfo{year}{1936}) \bibinfo{pages}{283--291}.
\bibitem[{Brush(1962)}]{Brush62}
\bibinfo{author}{S.~G. Brush},
\newblock \bibinfo{title}{Theories of liquid viscosity.},
\newblock \bibinfo{journal}{Chemical Reviews} \bibinfo{volume}{62}
  (\bibinfo{year}{1962}) \bibinfo{pages}{513--548}.
\bibitem[{Ewell and Eyring(1937)}]{Eyring37}
\bibinfo{author}{R.~H. Ewell}, \bibinfo{author}{H.~Eyring},
\newblock \bibinfo{title}{Theory of the viscosity of liquids as a function of
  temperature and pressure},
\newblock \bibinfo{journal}{The Journal of Chemical Physics}
  \bibinfo{volume}{5} (\bibinfo{year}{1937}) \bibinfo{pages}{726--736}.
\bibitem[{Babu and Sathian(2011)}]{Babu2011}
\bibinfo{author}{J.~S. Babu}, \bibinfo{author}{S.~P. Sathian},
\newblock \bibinfo{title}{The role of activation energy and reduced viscosity
  on the enhancement of water flow through carbon nanotubes},
\newblock \bibinfo{journal}{The Journal of Chemical Physics}
  \bibinfo{volume}{134} (\bibinfo{year}{2011}) \bibinfo{pages}{194509}.
\bibitem[{Bearman and Jones(1960)}]{Bearman60}
\bibinfo{author}{R.~J. Bearman}, \bibinfo{author}{P.~F. Jones},
\newblock \bibinfo{title}{Statistical mechanical theory of the viscosity
  coefficients of binary liquid solutions},
\newblock \bibinfo{journal}{The Journal of Chemical Physics}
  \bibinfo{volume}{33} (\bibinfo{year}{1960}) \bibinfo{pages}{1432--1438}.
\bibitem[{Tannehill et~al.(1997)Tannehill, Anderson, and
  Pletcher}]{Tannehill97}
\bibinfo{author}{J.~C. Tannehill}, \bibinfo{author}{D.~A. Anderson},
  \bibinfo{author}{R.~H. Pletcher}, \bibinfo{title}{Computational Fluid
  Mechanics Heat Transfer}, \bibinfo{publisher}{Taylor and Francis},
  \bibinfo{year}{1997}.
\bibitem[{{Alam} and {Bowman}(2002)}]{Alam2002}
\bibinfo{author}{J.~{Alam}}, \bibinfo{author}{J.~C. {Bowman}},
\newblock \bibinfo{title}{{Energy-Conserving Simulation of Incompressible
  Electro-Osmotic and Pressure-Driven Flow}},
\newblock \bibinfo{journal}{Theoretical and Computational Fluid Dynamics}
  \bibinfo{volume}{16} (\bibinfo{year}{2002}) \bibinfo{pages}{133--150}.
\bibitem[{Wesseling(2000)}]{Wesseling2000}
\bibinfo{author}{P.~Wesseling}, \bibinfo{title}{Principles of {C}omputational
  {F}luid {D}ynamics}, \bibinfo{publisher}{Springer}, \bibinfo{year}{2000}.
\bibitem[{Alam and Penney(2012)}]{Alam2012b}
\bibinfo{author}{J.~M. Alam}, \bibinfo{author}{J.~M. Penney},
\newblock \bibinfo{title}{A lagrangian approach for modelling electro-kinetic
  mass transfer in microchannels},
\newblock \bibinfo{journal}{International Journal of Heat and Mass Transfer}
  \bibinfo{volume}{55} (\bibinfo{year}{2012}) \bibinfo{pages}{7847 -- 7857}.
\bibitem[{Gozalpour et~al.(2005)Gozalpour, Ren, and Tohidi}]{Gozalpour2005}
\bibinfo{author}{F.~Gozalpour}, \bibinfo{author}{S.~R. Ren},
  \bibinfo{author}{B.~Tohidi},
\newblock \bibinfo{title}{{{CO}$_2$ EOR and Storage in Oil Reservoirs}},
\newblock \bibinfo{journal}{Oil \& Gas Science and Technology-Rev. IFP}
  \bibinfo{volume}{60(3)} (\bibinfo{year}{2005}) \bibinfo{pages}{537--546}.
\bibitem[{Roldán-Carrillo et~al.(2013)Roldán-Carrillo, Castorena-Cortés,
  Reyes-Avila, Zapata-Peñasco, and Olguín-Lora}]{Teresa2013}
\bibinfo{author}{T.~Roldán-Carrillo}, \bibinfo{author}{G.~Castorena-Cortés},
  \bibinfo{author}{J.~Reyes-Avila}, \bibinfo{author}{I.~Zapata-Peñasco},
  \bibinfo{author}{P.~Olguín-Lora},
\newblock \bibinfo{title}{Effect of porous media types on oil recovery by
  indigenous microorganisms from a mexican oil field},
\newblock \bibinfo{journal}{Journal of Chemical Technology \& Biotechnology}
  \bibinfo{volume}{88} (\bibinfo{year}{2013}) \bibinfo{pages}{1023--1029}.
\bibitem[{Afsharpoor et~al.(2012)Afsharpoor, Balhoff, Bonnecaze, and
  Huh}]{Afsharpoor2012}
\bibinfo{author}{A.~Afsharpoor}, \bibinfo{author}{M.~T. Balhoff},
  \bibinfo{author}{R.~Bonnecaze}, \bibinfo{author}{C.~Huh},
\newblock \bibinfo{title}{{CFD} modeling of the effect of polymer elasticity on
  residual oil saturation at the pore-scale},
\newblock \bibinfo{journal}{Journal of Petroleum Science and Engineering}
  \bibinfo{volume}{94~95} (\bibinfo{year}{2012}) \bibinfo{pages}{79 -- 88}.
\bibitem[{Koch and Brady(1985)}]{Koch85}
\bibinfo{author}{D.~L. Koch}, \bibinfo{author}{J.~F. Brady},
\newblock \bibinfo{title}{Dispersion in fixed beds},
\newblock \bibinfo{journal}{Journal of Fluid Mechanics} \bibinfo{volume}{154}
  (\bibinfo{year}{1985}) \bibinfo{pages}{399--427}.
\bibitem[{Hsu and Cheng(1990)}]{Hsu90}
\bibinfo{author}{C.~Hsu}, \bibinfo{author}{P.~Cheng},
\newblock \bibinfo{title}{Thermal dispersion in a porous medium},
\newblock \bibinfo{journal}{International Journal of Heat and Mass Transfer}
  \bibinfo{volume}{33} (\bibinfo{year}{1990}) \bibinfo{pages}{1587 -- 1597}.
\bibitem[{Gelhar et~al.(1992)Gelhar, Welty, and Rehfeldt}]{Gelhar92}
\bibinfo{author}{L.~W. Gelhar}, \bibinfo{author}{C.~Welty},
  \bibinfo{author}{K.~R. Rehfeldt},
\newblock \bibinfo{title}{A critical review of data on field-scale dispersion
  in aquifers},
\newblock \bibinfo{journal}{Water Resources Research} \bibinfo{volume}{28}
  (\bibinfo{year}{1992}) \bibinfo{pages}{1955--1974}.
\bibitem[{Wirasaet and Paolucci(2005)}]{Wirasaet2005}
\bibinfo{author}{D.~Wirasaet}, \bibinfo{author}{S.~Paolucci},
\newblock \bibinfo{title}{Adaptive wavelet method for incompressible flows in
  complex domains},
\newblock \bibinfo{journal}{Journal of Fluids Engineering}
  \bibinfo{volume}{127} (\bibinfo{year}{2005}) \bibinfo{pages}{656--665}.
\bibitem[{Alam et~al.(2006)Alam, Kevlahan, and Vasilyev}]{Alam2006}
\bibinfo{author}{J.~Alam}, \bibinfo{author}{N.~K.-R. Kevlahan},
  \bibinfo{author}{O.~Vasilyev},
\newblock \bibinfo{title}{Simultaneous space--time adaptive solution of
  nonlinear parabolic differential equations},
\newblock \bibinfo{journal}{Journal of Computational Physics}
  \bibinfo{volume}{214} (\bibinfo{year}{2006}) \bibinfo{pages}{829--857}.

\end{thebibliography}
